\documentclass[12pt,onecolumn]{article} % DRAFT

\usepackage{booktabs}
\usepackage{multirow}
\usepackage{amsfonts}
\usepackage{amssymb}
\usepackage{amsbsy} % Per a \boldsymbol
\usepackage{amsmath}
\usepackage[ansinew]{inputenc}%
\usepackage{cite}
\usepackage{multirow}
\usepackage{color}
\usepackage[usenames,dvipsnames]{xcolor}
\usepackage{lscape}
\usepackage{graphicx}
\usepackage{colortbl}
\usepackage{hyperref}
\usepackage[all]{xy}
\usepackage{algorithm}
\usepackage{algorithmic}

\def\X{\ensuremath{\mathbf{X}}}

\def\L{\ensuremath{\mathbf{L}}}

\def\W{\ensuremath{\mathbf{W}}}

\def\F{\ensuremath{\mathbf{F}}}

\def\x{\ensuremath{\mathbf{x}}}

\newcommand{\f}{{\mathbf f}}

 \def\mH{\mathcal{H}}

\definecolor{mt\_resi}{RGB}{255,102,51}
\definecolor{mt\_mea}{RGB}{0,235,50}
\definecolor{mt\_tr}{RGB}{51,155,0}
\definecolor{mt\_ro}{RGB}{120,120,120}
\definecolor{mt\_sh}{RGB}{51,255,255}
\definecolor{mt\_comm}{RGB}{255,0,0}
\definecolor{mt\_rail}{RGB}{204,204,204}
\definecolor{mt\_bare}{RGB}{153,102,51}
\definecolor{mt\_high}{RGB}{204,204,0}
 \definecolor{ma\_greyb}{RGB}{25, 25, 25}% 1 - grey buildings
 \definecolor{ma\_redb}{RGB}{255,0,0}% 2 -  red buildings
\definecolor{ma\_pool}{RGB}{25,225,255}% 3 - pools
\definecolor{ma\_asph}{RGB}{178,178,50}% 4 - asphalt
\definecolor{ma\_grass}{RGB}{0,100,0}% 5 - grass
 \definecolor{ma\_tree}{RGB}{0,225,0}% 6 - trees
\definecolor{ma\_wat}{RGB}{0,0,255}% 7 - water
\definecolor{ma\_conc}{RGB}{130,130,130}% 8 - concrete
\definecolor{ma\_tarmac}{RGB}{255,130,50}%9 - tarmac
\definecolor{ma\_whiteb}{RGB}{130,25,25}%10 - white buildings

\def\mH{\mathcal{H}}

\def\Real{\mathbb{R}}

\def\K{\ensuremath{\mathbf{K}}}
\def\L{\ensuremath{\mathbf{L}}}

\def\U{\ensuremath{\mathbf{U}}}

\def\W{\ensuremath{\mathbf{W}}}
\def\X{\ensuremath{\mathbf{X}}}

\def\aa{\boldsymbol{\alpha}}
\def\pphhii{\boldsymbol{\phi}}
\def\PHI{\boldsymbol{\Phi}}
\def\LL{\boldsymbol{\Lambda}}

\def\u{\ensuremath{\mathbf{u}}}

\def\u{\ensuremath{\mathbf{u}}}
\def\x{\ensuremath{\mathbf{x}}}

\newcommand{\green}[1]{\textcolor[rgb]{0,0.5,0}{#1}}

\newcommand{\diego}[2]{\textcolor[rgb]{0.1,0.5,0}{#2}}

\usepackage{hyperref}
\hypersetup{
    bookmarks=true,         % show bookmarks bar?
    unicode=false,          % non-Latin characters in Acrobat’s bookmarks
    pdftoolbar=true,        % show Acrobat’s toolbar?
    pdfmenubar=true,        % show Acrobat’s menu?
    pdffitwindow=false,     % window fit to page when opened
    pdfstartview={FitH},    % fits the width of the page to the window
    pdftitle={KEMA},    % title
    pdfauthor={Tuia},     % author
    pdfsubject={KEMA},   % subject of the document
    pdfcreator={Tuia},   % creator of the document
    pdfproducer={Tuia}, % producer of the document
    pdfkeywords={keyword1} {key2} {key3}, % list of keywords
    pdfnewwindow=true,      % links in new window
    colorlinks=true,       % false: boxed links; true: colored links
    linkcolor=blue,          % color of internal links (change box color with linkbordercolor)
    citecolor=blue,        % color of links to bibliography
    filecolor=blue,      % color of file links
    urlcolor=blue           % color of external links
}

\begin{document}

\title{Multi-temporal and multi-source remote sensing image classification by nonlinear relative normalization}

\date{}

\author{Devis Tuia, Diego Marcos, and Gustau Camps-Valls
\thanks{\bf Preprint. Paper published at ISPRS Journal of Photogrammetry and Remote Sensing 120, DOI: 10.1016/j.isprsjprs.2016.07.004}
\thanks{This work has been partly supported by the Swiss National Science Foundation (grant PZ00P2-136827, http://p3.snf.ch/project-136827), and the European Reseach Council (ERC) funding under the ERC-CoG-2014 SEDAL under grant agreement 647423.}
\thanks{DT and DM are  with the Department of Geography, University of Zurich, Switzerland. E-mail: devis.tuia@geo.uzh.ch, http://devis.tuia.googlepages.com, Phone: +41-44 635 52 60.
}
\thanks{GCV is with Image Processing Laboratory (IPL), Universitat de Val\`encia, Spain. E-mail: gustau.camps@uv.es, http://isp.uv.es. 
}}

\markboth{IEEE TGRS, Vol. XX, No. Y, Month Z 2016}{Tuia et al.: Nonlinear Alignment of Multi-temporal and Multi-source Remote Sensing Images}

\maketitle

\begin{abstract} 
Remote sensing image classification exploiting multiple sensors is a very challenging problem: data from different modalities are affected by spectral distortions and mis-alignments of all kinds, and this hampers re-using models built for one image to be used successfully in other scenes. In order to adapt and transfer models across image acquisitions, one must be able to cope with datasets that are not co-registered, acquired under different illumination and atmospheric conditions, by different sensors, and with scarce ground references. Traditionally, methods based on histogram matching have been used. However, they fail when densities have very different shapes or when there is no corresponding band to be matched between the images. An alternative builds upon \emph{manifold alignment}. Manifold alignment performs a multidimensional relative normalization of the data prior to product generation that can cope with data of different dimensionality (e.g. different number of bands) and possibly unpaired examples. 
Aligning data distributions is an appealing strategy, since it allows to provide data spaces that are more similar to each other, regardless of the subsequent use of the transformed data. In this paper, we study a methodology that aligns data from different domains in a nonlinear way through {\em kernelization}. We introduce the Kernel Manifold Alignment (KEMA) method, which provides a flexible and discriminative projection map,  exploits only a few labeled samples (or semantic ties) in each domain, and reduces to solving a generalized eigenvalue problem.  We successfully test KEMA in multi-temporal and multi-source very high resolution classification tasks, as well as on the task of making a model invariant to shadowing for hyperspectral imaging. 
\end{abstract}

%\begin{keywords}
Feature extraction, Manifold learning, Domain adaptation, Graph-based methods, Multi-temporal, Multi-source, Very high resolution, Classification, Kernel methods.
%\end{keywords}

\section{Introduction}\label{sec:intro}

Many real-life problems currently exploit heterogeneous sources of remote sensing data: forest ecosystems studies~\cite{Asn05,Asn06}, post-catastrophe assessment~\cite{Brun10,Tau11} or land-use updating~\cite{Bru01,Nie02,Amo13} take advantage of the wide coverage and short revisit time of remote sensing sensors. They typically design specific image processing pipelines to produce maps of a product of interest. Despite the promises of remote sensing to tackle such ambitious problems, two main obstacles prevent this technology from reaching a broader range of applications: on the one hand, there is generally a lack of labeled data present at each acquisition and, on the other hand, the models need to be capable of dealing with images obtained under different conditions and thus potentially with different sensors. 

Working under label scarcity has been extensively considered in recent remote sensing image processing literature by means of optimizing the use of the few available labels~\cite{Cam13}. 
In our view, the problem of adapting remote sensing classifiers boils down to compensating for a variety of distortions and mis-alignments: 
 for example, data resolution may differ or seasonal conditions might offer remarkable differences in the spectral signatures observed. When the images cover the same area, registration can be approximate. Moreover, each scene depends on its particular illumination and viewing geometry, which causes spectral signatures to shift among acquisitions~\cite{Mat14b}. As a consequence, it becomes difficult, often impossible, to re-use field data acquired on a given campaign to process newly acquired images. Transferring models from one remote sensing image acquisition to the other can be a very challenging task. 

Adapting classifiers to (even slightly) shifted  data distributions is an old problem in remote sensing, which started in the 1970s with the signature extension field~\cite{Fle75,Olt05}, and then evolved, due to the technological advances in both sensor and processing routines, into what is generally referred to as the {\em transfer learning} problem~\cite{Pan10,Pat14}. 
By transfer learning, we mean all kind of methodologies aiming at making models {\em transferable} across image/data acquisitions. In recent remote sensing literature, works have mainly considered three research directions~\cite{Tui15b}: 1) unifying the data representation, for example via atmospheric correction~\cite{Gua09}, feature selection~\cite{Bruzzone2009a}, or feature extraction~\cite{Vol14b,Sun15,Sun16}; 2) incorporating invariances in the classifier, for example via synthetic (`virtual') examples~\cite{Ver13} or physically-inspired features~\cite{Pac14,Ver10}; and 3) adapting the classifier to cope with the shift among acquisitions, for example via semi-supervised-inspired strategies~\cite{Raj06b,Bru10} or active learning~\cite{Mat12}.

Most of the methodologies above rely on the fact that all images are acquired by the same sensor (i.e. they share the same $d$-dimensional data space, as well as the nature -and physical meaning- of the features), or that all information and know-how necessary to convert to surface reflectance is available to the user performing the analysis, which is unfortunately often not the case. Moreover, at the application level there is generally no requirement of sticking to a specific sensor (taking the example of  post-catastrophe intervention,  the fact of waiting for the next cloud-free image of a specific sensor can mean the loss of human lives): since more and  more images are currently available to the general public and organizations, new transfer learning approaches must be capable to unify data from different sensors, at different resolutions, without co-registration, and without being specific to a given end classifier~\cite{Gom14}. The recently proposed \emph{manifold alignment} methods gather all these properties.

\begin{figure}
\includegraphics[width=\linewidth]{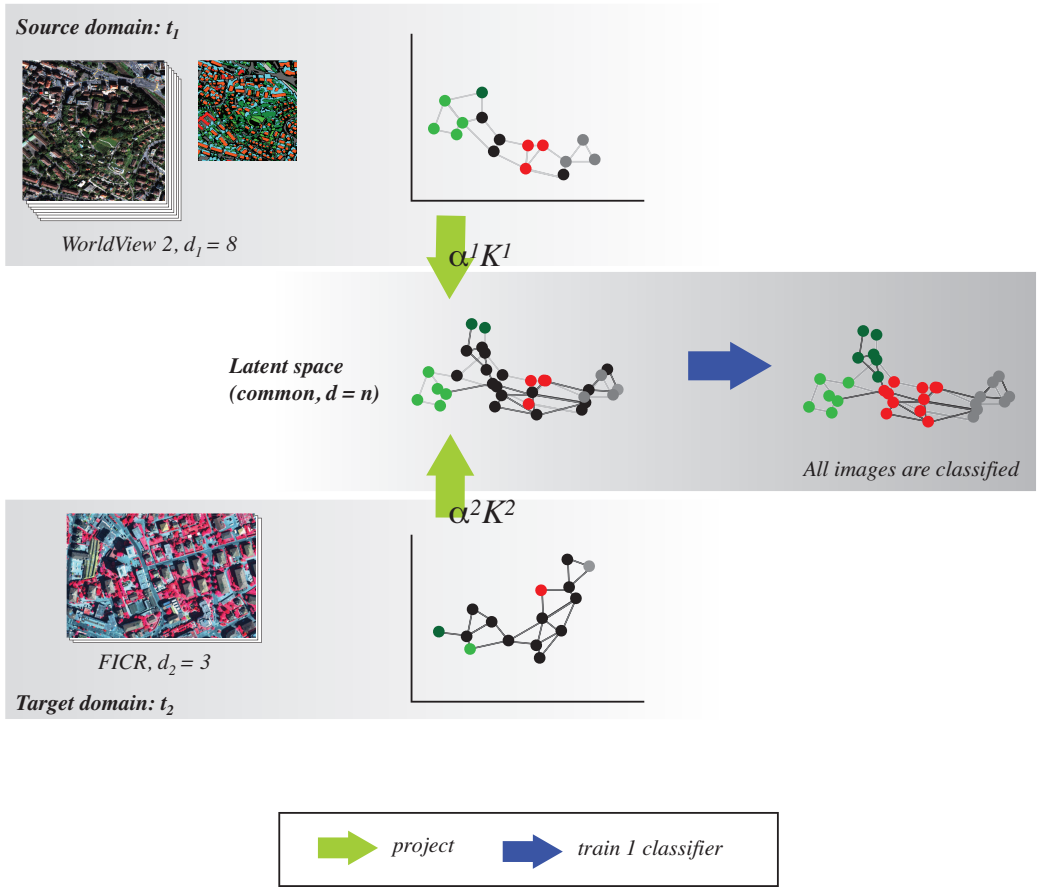}
\caption{Illustration of KEMA aligning data distributions in a multi-sensor setting.\label{fig:kema}}
\end{figure}

Manifold alignment~\cite{Wan11} is a machine learning framework aiming at matching, or \emph{aligning}, a set of domains (the images) of potentially different dimensionality using feature extraction under pairwise proximity constraints~\cite{Ham05}. In some sense, manifold alignment performs registration in the feature space and matches corresponding samples, where the correspondence is defined by a series of proximity graphs encoding some prior knowledge of interest (e.g. co-location, class consistency). An intuition of how manifold alignment functions is provided in Fig.~\ref{fig:kema}. Its application to remote sensing data is relatively recent: in~\cite{Tui13d}, authors presented the semi-supervised manifold alignment method (SSMA), which gathers all properties above, but at the price of requiring labeled pixels in all domains to perform the alignment. Authors in~\cite{Yan16} studies issued of spatial consistency and in~\cite{Yan16b} they propose a multi-scale alignment procedure not relying on labels in all domains. Finally, true colour visualization for hyperspectral data was tackled in~\cite{Lia16}.

In this paper, we study the effectiveness of the nonlinear counterpart of SSMA, the Kernel Manifold Alignment (KEMA~\cite{Tui15d}), as well as its relevance for remote sensing problems. KEMA is a flexible, scalable, and intuitive method for aligning manifolds. KEMA provides a flexible and discriminative projection function, only exploits a few labeled samples (or semantic ties~\cite{Mon13}, when images are roughly registered -- see Section~\ref{sec:wesma}) in each domain, and reduces to solving a simple generalized eigenvalue problem. 

KEMA is introduced in Section~\ref{sec:methodo}. In Section~\ref{sec:res}, we test it in several real-life scenarios, including multi-temporal and multi-source very high resolution image classification problems, as well as in the challenging task of making a model shadow-invariant in hyperspectral image classification. Section~\ref{sec:conc} concludes the paper.

\section{Kernel Manifold Alignment (KEMA)}\label{sec:methodo}
In this section, we detail the KEMA method. We first recall the linear counterpart, the SSMA method~\cite{Wan11b}. Noting the main problems of this method, we introduce KEMA as a solution to address them. The reader interested in more theoretical details of KEMA can find them in~\cite{Tui15d}. Code can be found at the URL: \texttt{\url{https://github.com/dtuia/KEMA}}. 

\subsection{Notation}

To fix notation, we consider a series of $M$ domains. For each one of them, we have a data set: ${\mathcal M} := \{\x_i^m\in \mathbb{R}^{d_m}|i=1,\ldots,{n_m} \}$, where $n_m$ is the number of samples issued from domain $m$ with data dimensionality $d_m$, and $m=1,\ldots,M$. Some of the pixels in $\x_i$ are labeled ($l_1, ..., l_M$), and most are unlabeled. From one domain to another, the data are not necessarily semantically paired, i.e. $n_1 \neq n_m \neq n_M$, nor it is mandatory that all domains have the same dimension, i.e. $d_1 \neq d_m \neq  d_M$. 

\subsection{Semi-supervised manifold alignment (SSMA)}

The linear SSMA method was originally proposed in~\cite{Wan11b} and successfully adapted to remote sensing problems in~\cite{Tui13d}. The SSMA method aligns data from all $M$ domains by projecting them into a common \emph{latent space} using a set of domain-specific projection functions, $\f^m$, collectively grouped into the projection matrix ${\bf F}:=[{\bf f}^1, \ldots, {\bf f}^M]^\top$. The latent space has two properties: it is discriminant for classification and respects the original geometry of each manifold. To do so, SSMA tries to find a data projection matrix ${\bf F}$ that maximizes the following cost function
$$\mathcal{L} = \dfrac{\mu \texttt{GEO} + \texttt{SIM}}{\texttt{DIS}},$$ 
where we aim to maximize a topology/geometry (\texttt{GEO}) and a class similarity (\texttt{SIM}) terms while minimizing a class dissimilarity term (\texttt{DIS}) between all samples, and $\mu>0$ is a parameter controlling the contribution of the similarity and the topology terms. The three terms correspond to: 
\begin{enumerate}
\item a geometry-preservation term, \texttt{GEO}, forcing the local geometry of each manifold to remain unchanged, i.e. penalizing projections mapping neighbors in the input space far from each other,
\begin{align}\label{g}
\texttt{GEO} &= \sum_{m=1}^M \sum_{i,j=1}^{{n_{m}}} W_g^{m}(i,j) {\|{\f^{m}}^\top{\x_i^m} - {\f^{m}}^\top{\x_j^{m}} \|^2}\nonumber\\ &  = \text{tr}(\F^\top\X\L_g\X^\top\F),
\end{align}
where $W_g^m$ is a similarity matrix returning the value $1$ if two pixels of domain $m$ are neighbours in the original feature space and $0$ otherwise. $W_g^m$ is typically a $k$-NN graph. $\L_g$ is the $(\sum_m n_m \times \sum_m n_m)$ graph Laplacian issued from the similarity matrices $\W_g^m$, stacked in a block-diagonal matrix. All the out-of-diagonal blocks of $\W_g$ are empty, since we do not want to preserve neighbourhood relationships between the images.

\item a class similarity term, \texttt{SIM}, penalizing projections mapping samples of the same class far from each other, 
\begin{align}\label{s}
\texttt{SIM} &= \sum_{m,m'=1}^M\sum_{i,j=1}^{l_{m},l_{m'}} W_s^{m,m'}(i,j) {\| {\f^{m\top}}\x_i^m - {\f^{m'\top}}\x_j^{m'} \|^2} \nonumber\\ & = \text{tr}(\F^\top\X\L_s\X^\top\F),
\end{align}
where $W_s^{m,m'}$ is a similarity matrix returning the value $1$ if two pixels from domains $m$ and $m'$ belong to the same class. These are the tie points performing registration in the spectral space, and are used to match the images to each other.

\item a class dissimilarity term, \texttt{DIS}, penalizing projections mapping pixels of different classes close to each other. 
\begin{align}\label{d}
\texttt{DIS} &= \sum_{m,m'=1}^M\sum_{i,j=1}^{l_{m},l_{m'}} W_d^{m,m'}(i,j) {\| {\f^{m\top}}\x_i^m - {\f^{m'\top}}\x_j^{m'} \|^2} \nonumber\\ & = \text{tr}(\F^\top\X\L_d\X^\top\F),
\end{align}
where $W_d^{m,m'}$ is a dissimilarity matrix returning the value $1$ if two pixels from domains $m$ and $m'$ belong to different classes. These tie points prevent the solution to collapse in a single point and, together with the \texttt{SIM} term, foster the latent space to be discriminative.
\end{enumerate}
Now, by combining Eqs.~\eqref{g}-\eqref{d}, it is straightforward to show that the solution boils down to finding the last eigenvalues of the following generalized eigenproblem~\cite{Wan11b}, which is directly derived: 
\begin{equation}
\X(\mu \L_g + \L_s)\X^\top \boldsymbol\varphi = \lambda \X \L_d \X^\top \boldsymbol\varphi,
\label{eq:ssma}
\end{equation}
where ${\bf X}$ is a $(d \times \sum_m n_m)$ block-diagonal matrix containing the data from the different domains to be aligned. $ \boldsymbol\varphi$ is the researched common projection matrix of size $d\times d$, with $d = \sum_{m=1}^M d_m$. 
The rows of $\boldsymbol\varphi$ contain a block of projectors for each domain, scaled by $\lambda^{1/2}$, in a particular block structure:
\begin{equation}
\F = \lambda^{\frac{1}{2}} \boldsymbol\varphi = 
\begin{bmatrix} 
\green{\f^1}\\ \vdots \\ \f^M  
\end{bmatrix}  = 
\begin{bmatrix} 
\green{f_{1,1}} & \green{\ldots} & \green{ f_{1,d}}\\
\green{\vdots} & \green{\ddots} & \green{\vdots}\\
\green{f_{d_1,1}} & \green{\ldots} & \green{f_{d_1,d}}\\
f_{d_1+1,1} & \ldots & f_{d_1+1,d}\\
\vdots & \ddots & \vdots\\
f_{d,1} & \ldots & f_{d,d}\\
\end{bmatrix},
\end{equation} 
where the eigenvectors for the first domain are highlighted in green. 

Once the projection matrix $\boldsymbol\varphi$ is obtained, any sample $\x_i^m\in\Real^{d_m \times 1}$ from domain $m$ (one of the domains considered) can be projected in the latent space by using the corresponding ($d_m \times d$) block of eigenvectors $\f^m$:
\begin{equation}
{\mathcal P}(\x_i^m) = \f^{m\top}\x_i^m.
\end{equation}
As for (k)PCA and other methods based on eigen-decomposition, the data can be projected onto a subspace of dimension $p$ lower than $d$ by simply using only the first $p\ll d$ columns of $\f^m$. In this sense, SSMA leaves some control on the dimensionality of the latent space for class separation.

\subsection{Kernel Manifold Alignment (KEMA)}

The idea behind {\em kernelization} is to map the data into a high dimensional Hilbert space $\mH$ with the mapping function $\pphhii(\cdot):\x\mapsto\pphhii(\x)\in \mH$  such that the mapped data is better suited for solving our problem. This technique has found wide adoption in many remote sensing data analysis problems~\cite{CampsValls09wiley}. In practice, computing this mapping explicitly can be prohibitive due to its high dimensionality. This can be avoided by expressing the problem in terms of dot products within $\mH$. We can then define an easy-to-compute kernel function $k(\x_i,\x_j) = \langle \pphhii(\x_i),\pphhii(\x_j)\rangle_{\mH}$ returning similarities between mapped samples without having to compute $\pphhii(\cdot)$ explicitly.

In the multi-modal setting considered here, we would have to map the $M$ datasets to $M$ Hilbert spaces $\mH_m$ of dimension $H_m$, $\pphhii_m(\cdot):\x\mapsto\pphhii_m(\x)\in \mH_m$, $m=1,\ldots,M$. Then, we replace all the samples with their mapped feature vectors. The \texttt{GEO}, \texttt{SIM} and \texttt{DIS} terms become:

{\small 
\begin{align}\label{gg}
\texttt{GEO} &= \sum_{m=1}^M \sum_{i,j=1}^{{n_{m}}} W_g^{m}(i,j) {\|{\u^{m}}^\top{\pphhii(\x_i)^m} - {\u^{m}}^\top{\pphhii(\x_j)^{m}} \|^2}\nonumber\\ &  = \text{tr}(\U^\top\PHI\L_g\PHI^\top\U)
\\
\texttt{SIM} &= \sum_{m,m'=1}^M\sum_{i,j=1}^{l_{m},l_{m'}} W_s^{m,m'}(i,j) {\| {\u^{m\top}}\pphhii(\x_i)^m - {\u^{m'\top}}\pphhii(\x_j)^{m'} \|^2} \nonumber\\ & = \text{tr}(\U^\top\PHI\L_s\PHI^\top\U),
\\
\texttt{DIS} &= \sum_{m,m'=1}^M\sum_{i,j=1}^{l_{m},l_{m'}} W_d^{m,m'}(i,j) {\| {\u^{m\top}}\pphhii(\x_i)^m - {\u^{m'\top}}\pphhii(\x_j)^{m'} \|^2} \nonumber\\ & = \text{tr}(\U^\top\PHI\L_d\PHI^\top\U),\label{dd}
\end{align}
}

As for the SSMA case, combining Eqs.~\eqref{gg}-\eqref{dd} leads to a generalized eigendecomposition problem:
$$\PHI (\L_g + \mu \L_s)\PHI^\top \U = \lambda \PHI \L_d \PHI^\top \U,$$
where $\PHI$ is a block diagonal matrix containing the data matrices $\PHI^m=[\pphhii_m(\x_1), \ldots, \pphhii_m(\x_{n_m})]^\top$ and $\U$ contains the eigenvectors organized in rows for the particular domain defined in Hilbert space $\mH_m$, $\U =[\u_1, \u_2, \ldots, \u_H]^\top$ where $H=\sum_{m=1}^M H_m$.  
As stressed above, $\PHI$ and $\U$ live in a high dimensional space that might be very costly or even impossible to compute. Therefore, we express the eigenvectors as a linear combination of mapped samples using the Representer's theorem~\cite{Yan07} , $\u^m=\PHI^m\aa^m$ (or $\U =\PHI\LL$ in matrix notation):
\begin{equation}\label{primalKEMA}
\K (\L _g+ \mu \L_s)\K\LL = \lambda \K \L_d \K \LL,
\end{equation}
where $\K$ is a block diagonal matrix containing the kernel matrices $\K^m$. Now the eigenproblem becomes of size $n\times n$ instead of $d\times d$, and we can extract a maximum of $n$ components.  
\begin{equation}
\LL = 
\begin{bmatrix} 
\green{\aa^1}\\ \vdots \\ \aa^M  
\end{bmatrix}  = 
\begin{bmatrix} 
\green{\alpha_{1,1}} & \green{\ldots} & \green{ \alpha_{1,n}}\\
\green{\vdots} & \green{\ddots} & \green{\vdots}\\
\green{\alpha_{n_1,1}} & \green{\ldots} & \green{\alpha_{n_1,n}}\\
\alpha_{n_1+1,1} & \ldots & f_{\alpha_1+1,n}\\
\vdots & \ddots & \vdots\\
\alpha_{n,1} & \ldots & \alpha_{n,n}\\
\end{bmatrix} .
\end{equation} 

This dual formulation is advantageous when dealing with very high dimensional datasets, $d\gg n$ for which the SSMA problem is not well-conditioned. Operating in $Q$-mode endorses the method with numerical stability and computational efficiency in current high-dimensional problems, e.g. when using Fisher vectors or deep features for data representation. {This type of problems with much more dimensions than points are becoming more and more prominent in remote sensing~\cite{Lag15,Mar15}. In this sense, even KEMA with a linear kernel (which corresponds to the SSMA solution) becomes a valid solution for these problems, as it has all the advantages of methods related to (kernel) Canonical Correlation Analysis ((k)CCA~\cite{Lai00}), but can also deal with unpaired data.}

Projection of a new test vector $\x_i^m$ to the latent space requires first mapping it to its corresponding kernel form $\K_i^m$ and then applying the corresponding projection vector $\aa^{m}$ defined therein:
\begin{equation}
{\mathcal P}(\x_i^m) = \u^{m\top}\PHI_i^m = \aa^{m\top}\PHI^{m\top}\PHI_i^m = \aa^{m\top}\K_i^m,
\label{eq:proj}
\end{equation}
where $\K_i^m$ is a vector of kernel evaluations between sample $\x_i$ and all samples from domain $m$ used to define the projections $\aa^{m}$. 
Therefore, projection to the kernel latent space is possible through the use of dedicated reproducing kernel functions. %

\section{Experimental Results}\label{sec:res}
In this section, we present experimental results in three challenging remote sensing problems: multi-temporal /  multi-source VHR classification, shadow removal in hyperspectral images, and multi-source image alignment without labels.

\subsection{Multi-temporal and multi-sensor VHR classification}
The first experiment is a direct comparison to the multi-source experiment reported in~\cite{Tui13d}. We consider three VHR images (Fig.~\ref{fig:im}) depicting peri-urban settlements:

\begin{itemize}
\item[-] \emph{Prilly}: the first image is acquired by the WorldView-2 VHR satellite (8 visible and near-infrared bands) over Prilly, a residential neighborhood of Lausanne, Switzerland. The image is acquired on August 2, 2011 and has been pansharpened using the Gram-Schmid transform to a resolution of approximatively 0.7m.
\item[-] \emph{Malley}: the second image is also acquired by WorldView-2  over another residential neighborhood of Lausanne, Montelly. The image is acquired on September 29, 2010 and has also been pansharpened using the Gram-Schmid transform to 0.7m.
\item[-] \emph{Zurich}: the third image is acquired by the QuickBird satellite (4 bands, RGB- NIR) over a residential neighborhood of Zurich, Switzerland. The image has been acquired on October 6, 2006 and pansharpened.
\end{itemize}

\definecolor{resi}{RGB}{255,102,51}
\definecolor{mea}{RGB}{0,235,50}
\definecolor{tr}{RGB}{51,155,0}
\definecolor{ro}{RGB}{120,120,120}
\definecolor{sh}{RGB}{51,255,255}
\definecolor{comm}{RGB}{255,0,0}
\definecolor{rail}{RGB}{160,160,160}
\definecolor{bare}{RGB}{153,102,51}
\definecolor{high}{RGB}{204,204,0}

\begin{figure}[!b]
\centering
\setlength{\tabcolsep}{2pt}
\begin{tabular}{ccc}
\includegraphics[width=.3\linewidth]{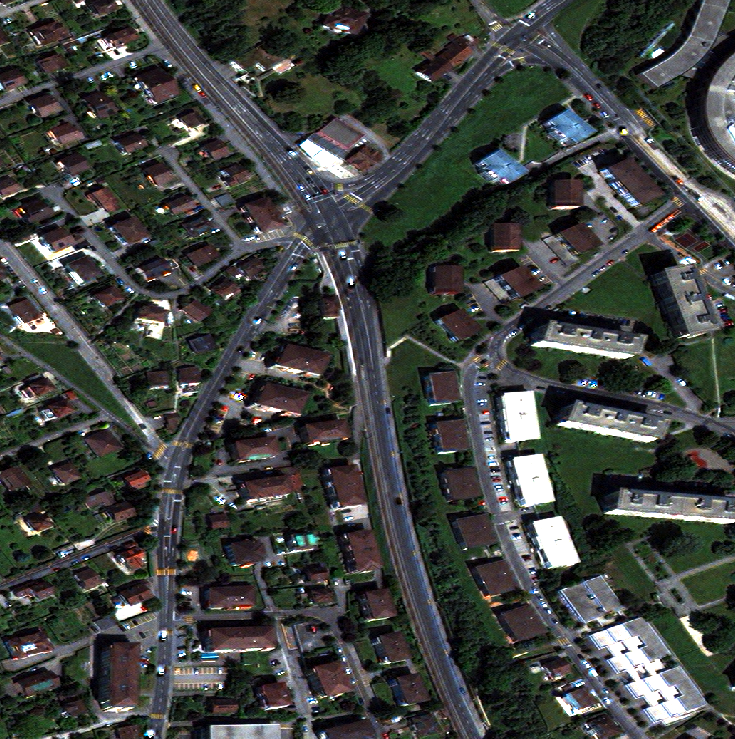}&
\includegraphics[width=.3\linewidth]{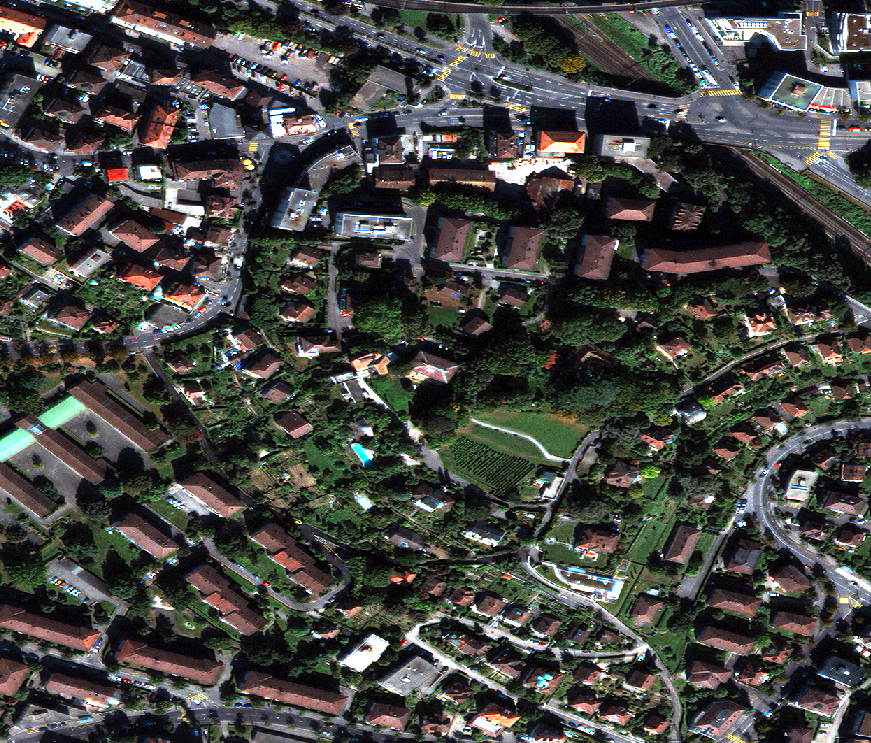}&
\includegraphics[width=.3\linewidth]{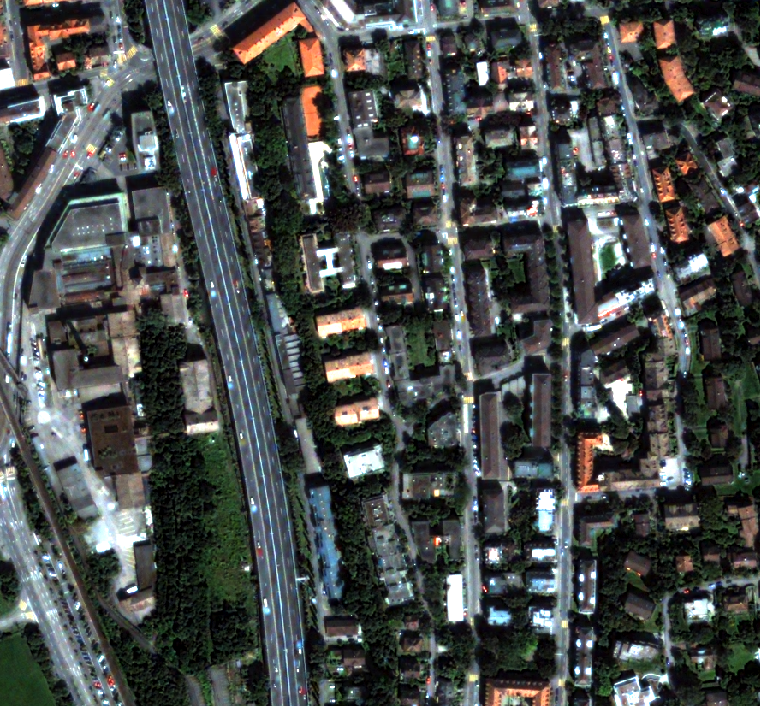}\\
%\multicolumn{3}{c}{(A) Images} \\ 
\includegraphics[width=.3\linewidth]{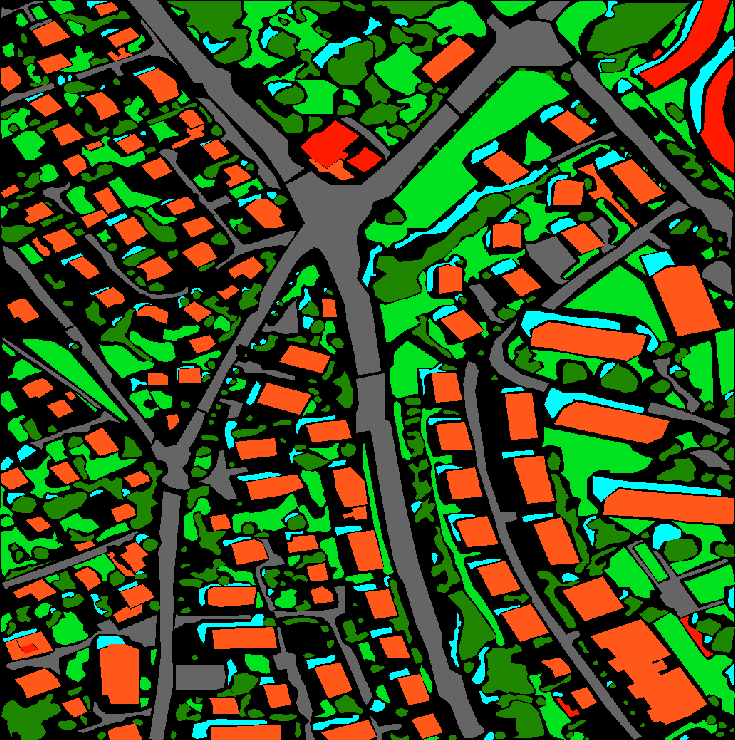}&
\includegraphics[width=.3\linewidth]{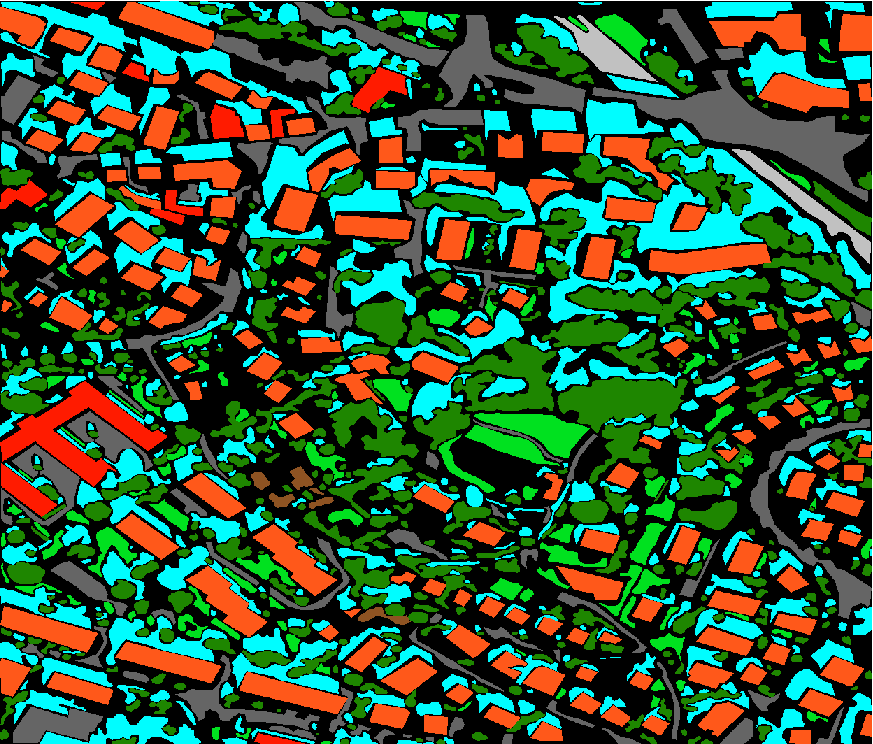}&
\includegraphics[width=.3\linewidth]{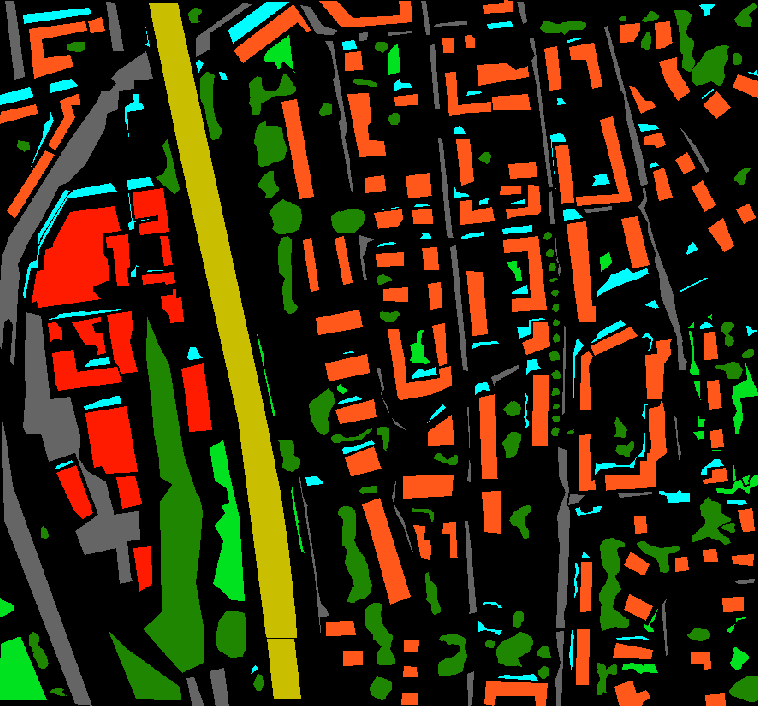}\\
Prilly (WV2) & Montelly (WV2) &Zurich (QB) \\
%\multicolumn{3}{c}{(B) Classes} \\
\end{tabular}
\caption{The WorldView-2 (WV2) and QuickBird (QB) images used in the remote sensing semantic classification experiments. {Color legend: \textcolor{resi}{residential}, \textcolor{mea}{meadows}, \textcolor{tr}{trees}, \textcolor{ro}{roads}, \textcolor{sh}{shadows}, \textcolor{comm}{commercial building}, \textcolor{rail}{railway}, \textcolor{bare}{bare soil}, \textcolor{high}{highway}.}}
\label{fig:im}
\end{figure}

\begin{figure*}[!t]
\begin{tabular}{llccc}
\setlength{\tabcolsep}{1pt}

&& \multicolumn{3}{c}{Image predicted} \\
&& Prilly & Montelly & Zurich (4 bands) \\
\multirow{3}{*}{\rotatebox{90}{\hspace{-0.5cm} Leading training image}}&

\rotatebox{90}{Prilly} & 
\includegraphics[width=5.2cm]{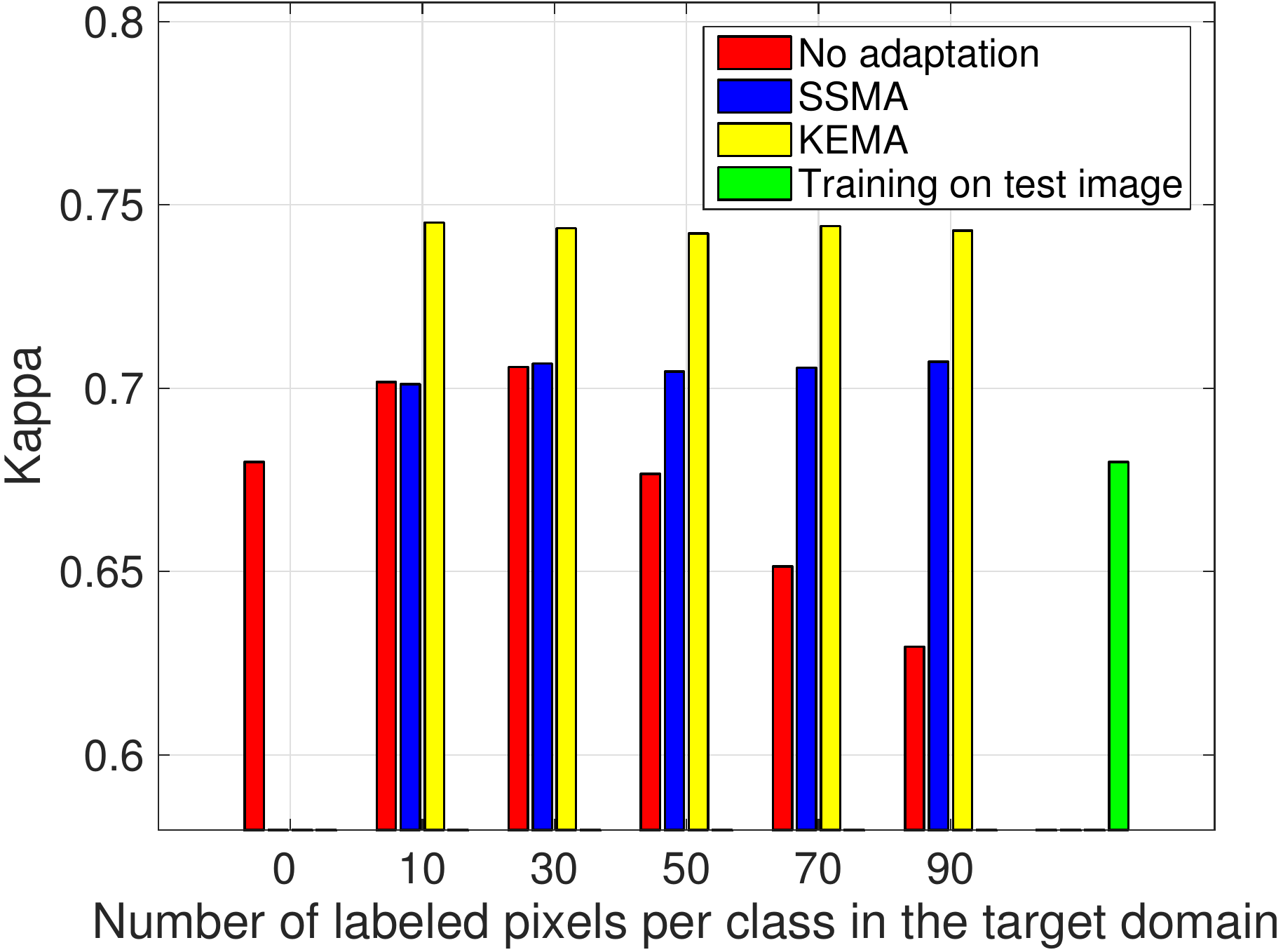}&
\includegraphics[width=5.2cm]{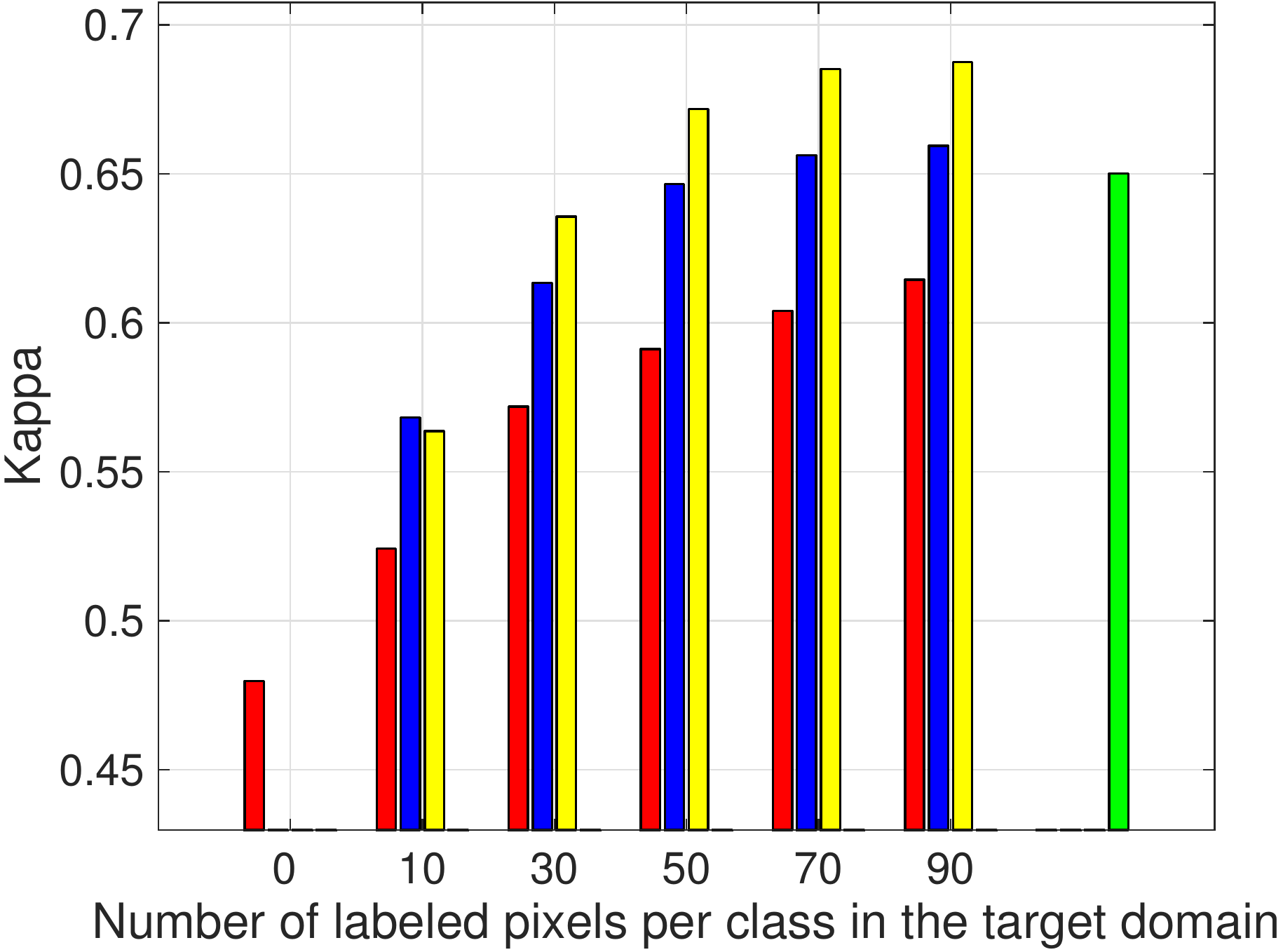} &
\includegraphics[width=5.2cm]{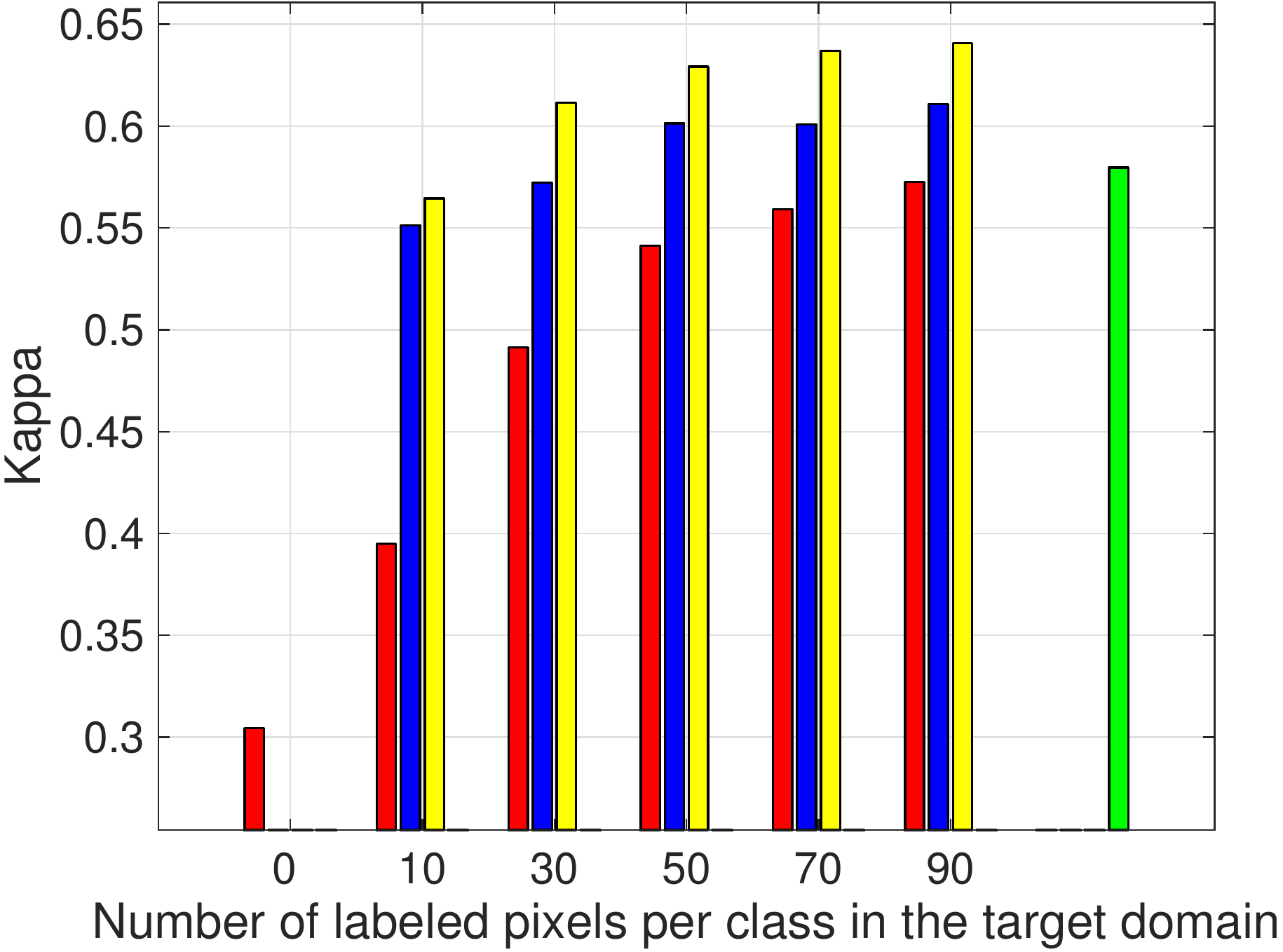}  \\
& \rotatebox{90}{Montelly} & 
\includegraphics[width=5.2cm]{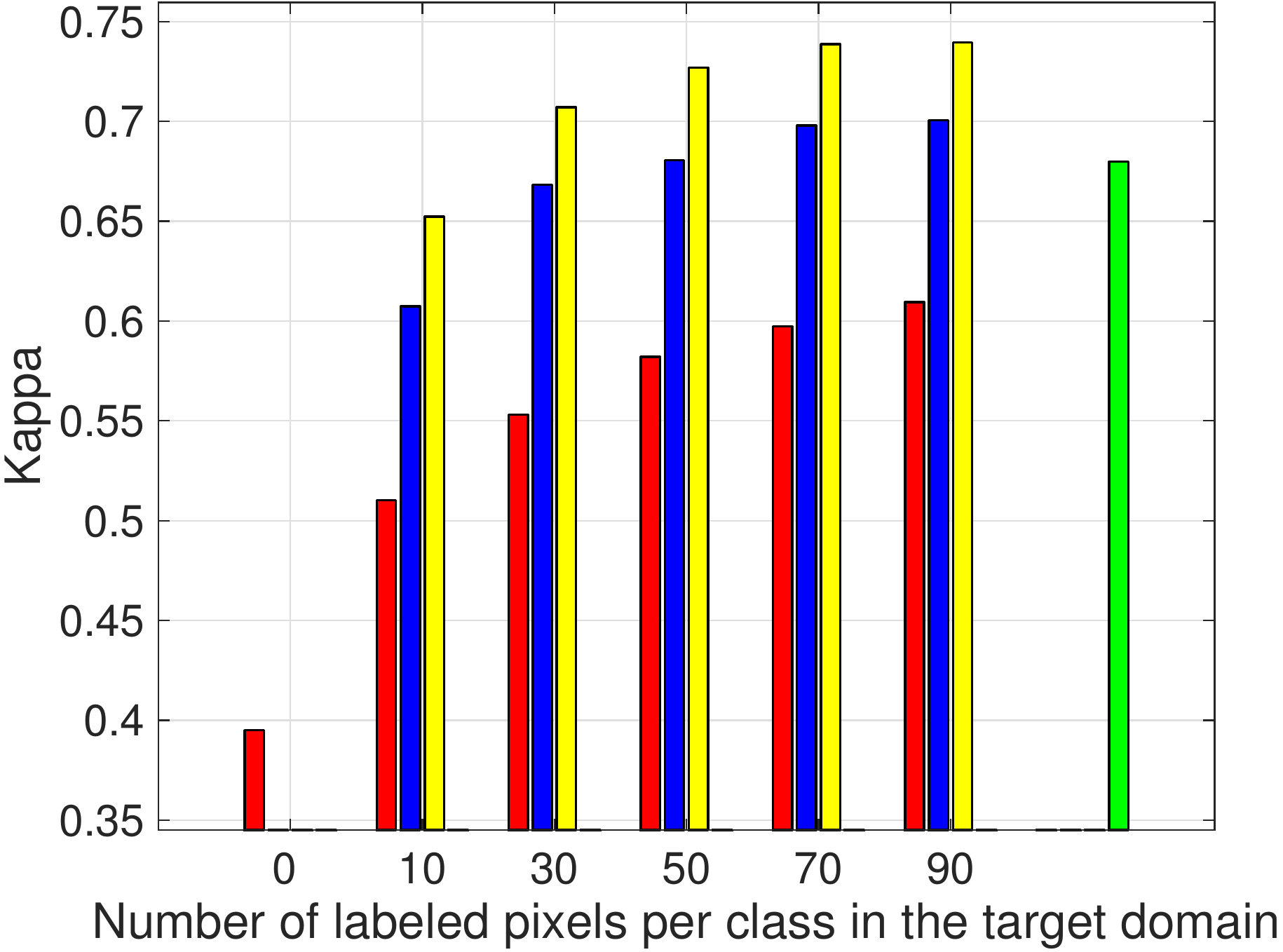}&
\includegraphics[width=5.2cm]{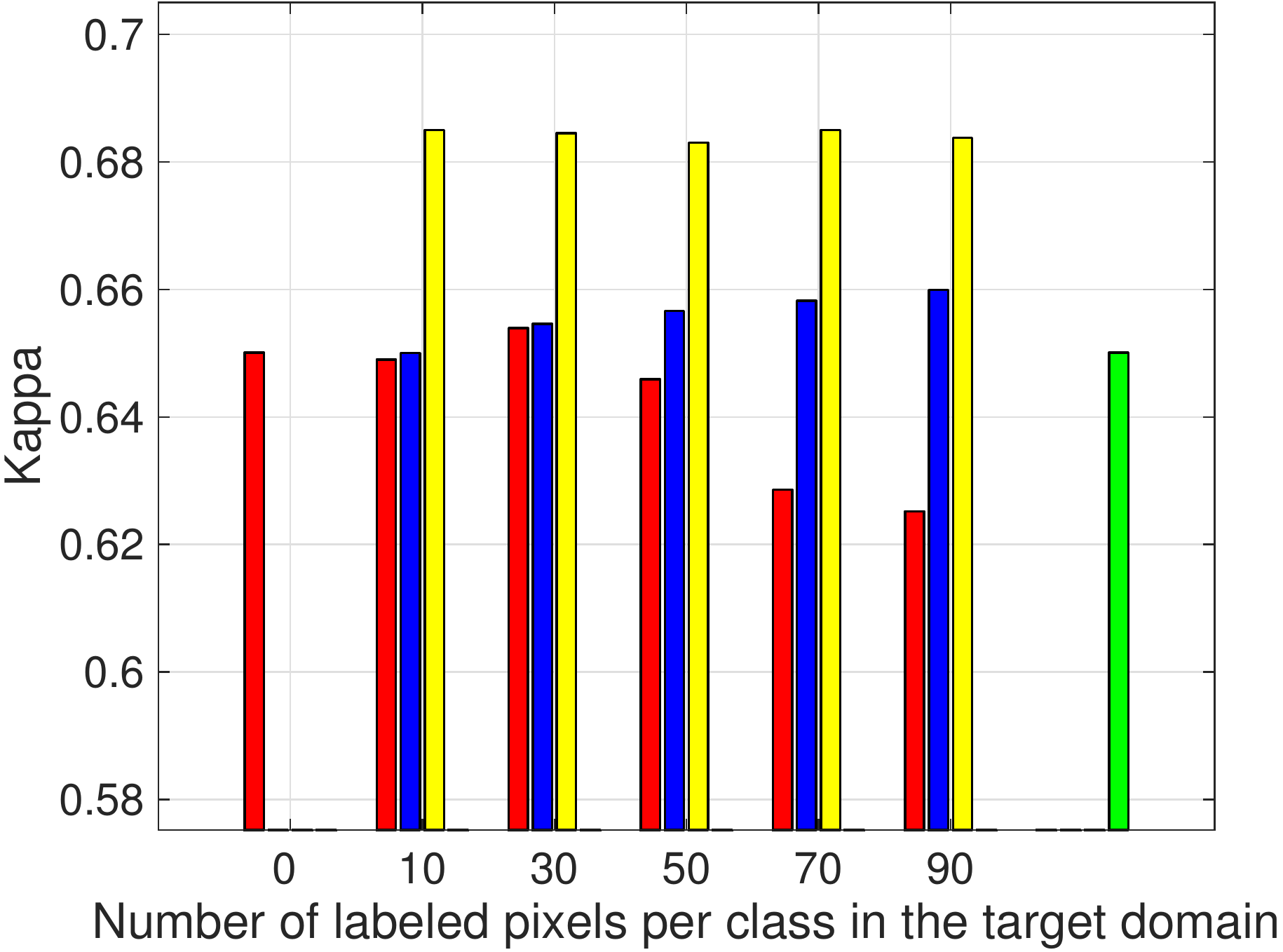}&
\includegraphics[width=5.2cm]{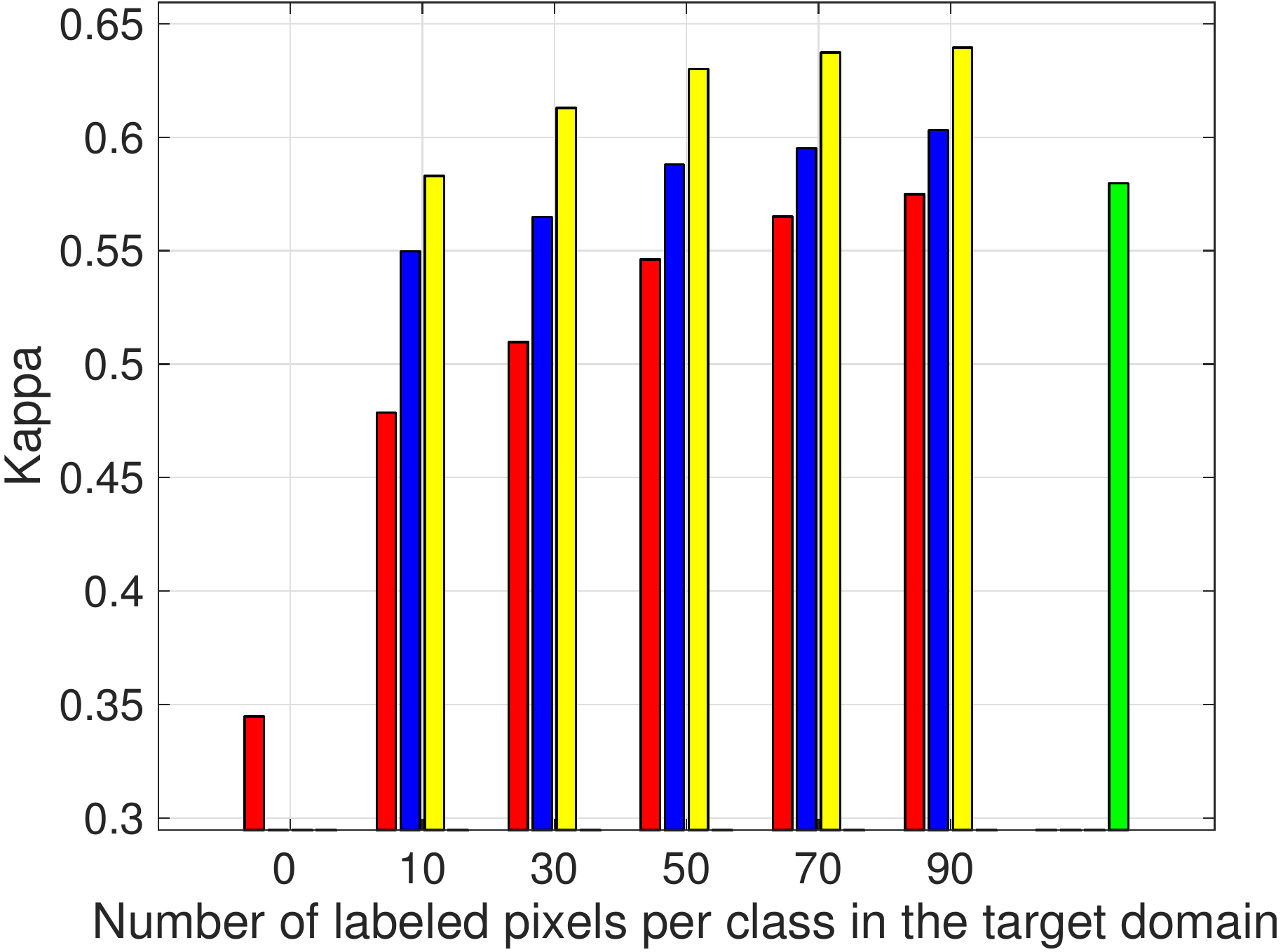}\\

& \rotatebox{90}{Zurich (QuickBird, 4 bands)} &
\includegraphics[width=5.2cm]{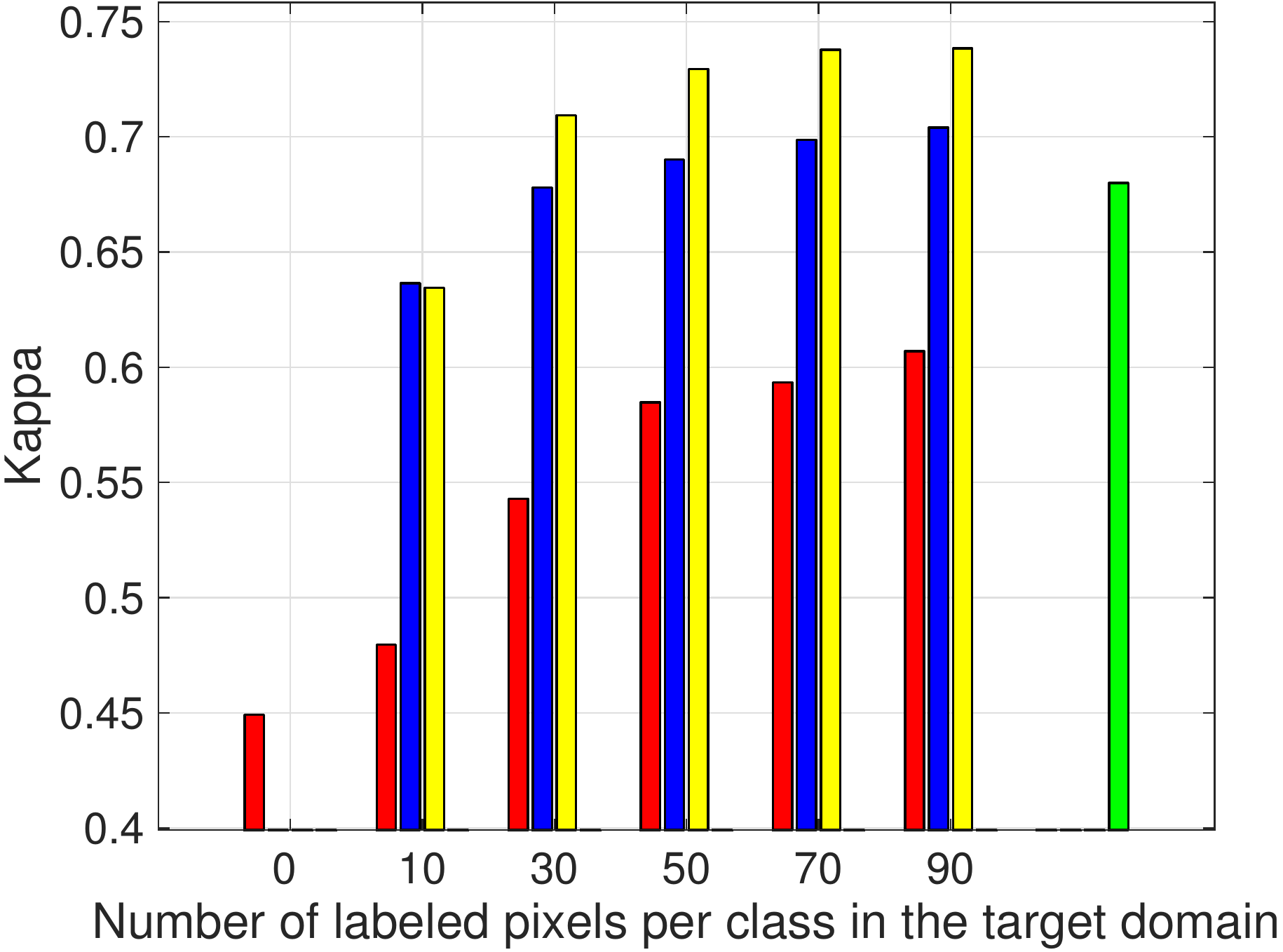}&
\includegraphics[width=5.2cm]{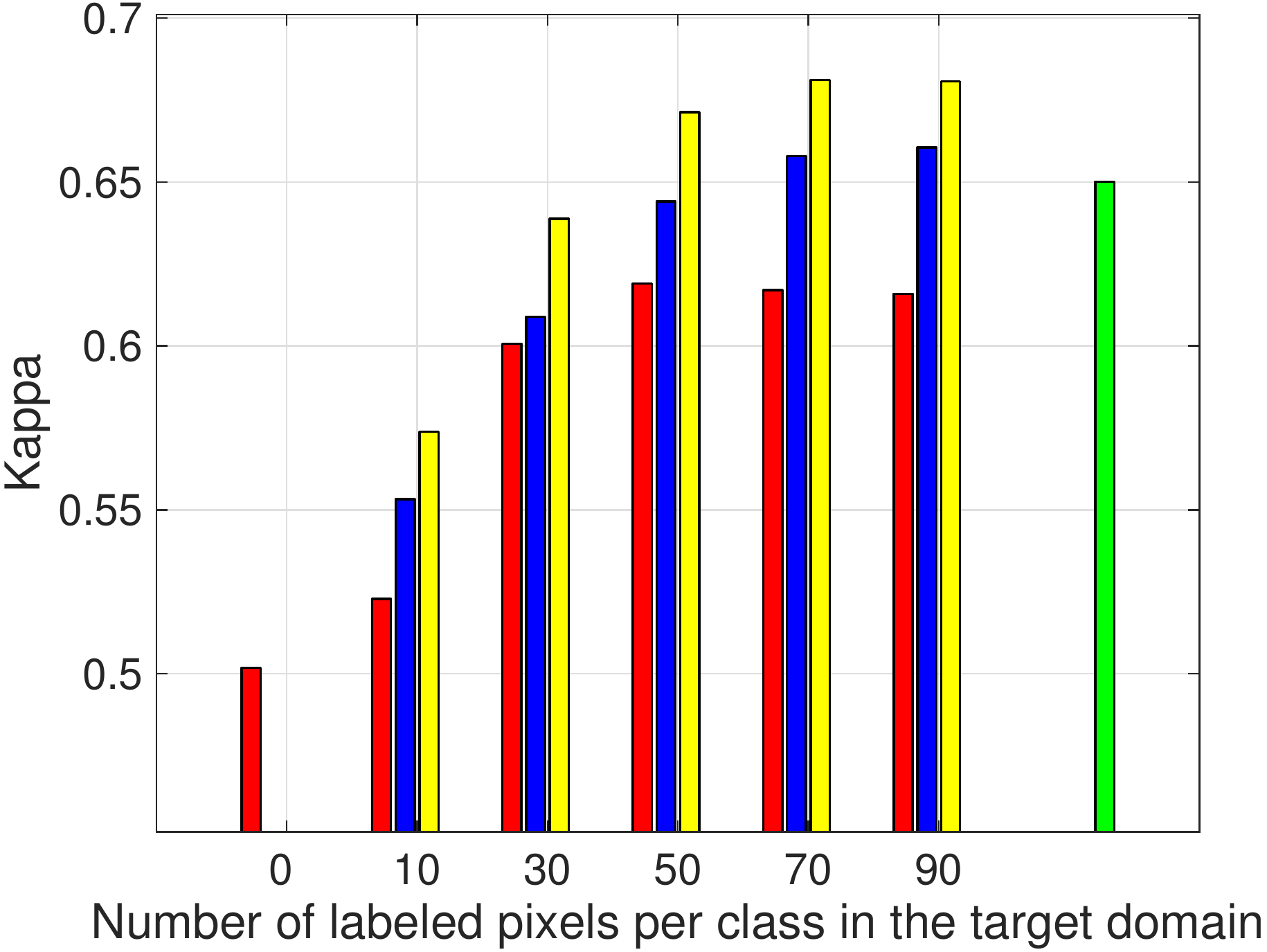}&
\includegraphics[width=5.2cm]{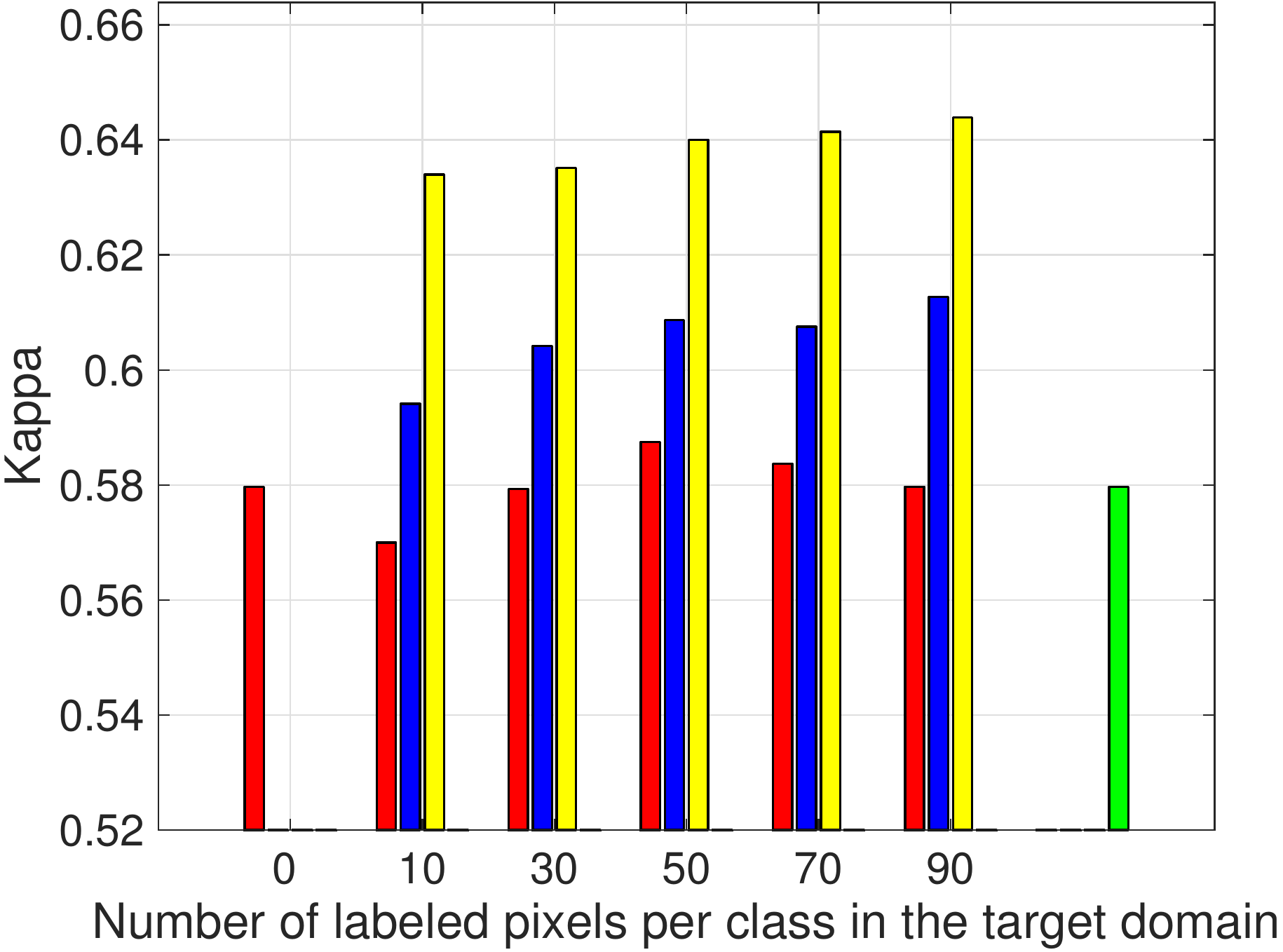}\\
\end{tabular}

\caption{Numerical results for the multi-source experiment. Rows indicate the image from which 100 labeled pixels {\em per} class are used ($l_1 = 100$ {\em per} class). $\kappa$ performances for increasing number of labeled pixels in the two other images ($l_{2} = l_3 = [10, ..., 90]$ {\em per} class) are reported. Columns correspond to the image that has been used for testing. The baseline is the model obtained using 100 pixels {\em per} class from the test image only.}
\label{fig:3domainsWVQBdiag}

\end{figure*}

For each image, a ground truth consisting of 9 classes is available (see bottom row of Fig.~\ref{fig:im}). We follow the experimental protocol of~\cite{Tui13d}: from all the available labeled pixels in each image, 50\% are kept apart as the testing set. The remaining 50\% are used to extract the labeled and unlabeled pixels composing the $\x_m$ sets. We then extract $l_1=100$ labeled pixels per class from what we call the leading domain image, which is the image carrying most labeled samples (we take each image in turn as the leading domain image). 
In our setting, we also need labeled pixels from the two other acquisitions: we tested an increasing additional of labeled samples, $l_2 = l_3 = [10, 30, 50, 90]$ pixels per class. As in~\cite{Tui13d}, the unlabeled examples are selected using an iterative clustering algorithm, the bisecting $k$-means~\cite{Kas09}, which runs $k$-means with $2$ clusters iteratively, by splitting the current largest cluster in the dataset. This way, we sample  $500$ unlabeled examples per each image source.  We  use the labeled and unlabeled examples to extract both the SSMA and KEMA projections and then project all images in the latent space. Finally, we use all the projected labeled examples to train a single classifier (a linear SVM) in the latent space. This classifier is used to predict all the test pixels of all three images at once (i.e. no specific training is performed for the specific images separately).

In KEMA, we use RBF kernels with the bandwidth $\sigma_{m}$ fixed as half the median distance between the samples of the specific image (labeled or unlabeled). By doing so, we allow different kernels in each domain, thus tailoring the similarity function to the data structure observed~\cite{Tui15d}. To build the graph Laplacians, we used a series of graphs built using $k$-NN graphs with $k = 9$ as in~\cite{Tui13d}. We validated the optimal number of dimensions, as well as the optimal $C$ parameter in the SVM classifier using the labeled samples in a cross-validation setting. Finally, as in~\cite{Tui13d} we add a baseline, which is the classifier learned with the original features. Since the Zurich image has a different input space than the two others, only the common bands between QuickBird and WorldView-2 are considered.

The results are reported in Fig.~\ref{fig:3domainsWVQBdiag}. Two distinct behaviours are observed: 
\begin{itemize}
\item[-] Diagonal blocks of Fig.~\ref{fig:3domainsWVQBdiag} (when predicting the leading domain image, which carried most labels): in this cases, the predictions of KEMA are better than those of SSMA by $\approx 2-5\%$ and remain consistent  when adding samples from the other domains. This means that the images are aligned correctly and the inclusion of labels from other images does not disturb the classifier (as in the `no adaptation' case). On the contrary, adding labeled samples from the other images is beneficial, as one can observe by comparing the KEMA results with the optimal case obtained when using only the 100 labeled pixels per class from the leading image (green bars): the final prediction is 5-10\% more accurate than in the case, where the leading image is used alone (i.e. without extra labeled samples coming from the other acquisitions). This means that the extra labeled are aligned correctly, since the classifier trained with $100+l_2+l_3$ aligned examples per class outperforms the one obtained with $100$ pixels per class.
\item[-] Off-diagonal blocks of Fig.~\ref{fig:3domainsWVQBdiag} (when predicting the two other, scarcely labeled images): in the off-diagonal blocks we can observe  a constant improvement of the results obtained by SSMA, which corresponds already to a strong improvement over the `no adaptation' case. The improvement of KEMA with respect to the latter is more striking ($\approx 5-15\%$) when using little labels from the test images. In comparison to SSMA we observe a constant $3-5\%$ improvement.
\end{itemize}

\subsection{Shadow compensation in hyperspectral image classification}
\begin{figure}[!t]
(a) \includegraphics[width=.95\linewidth]{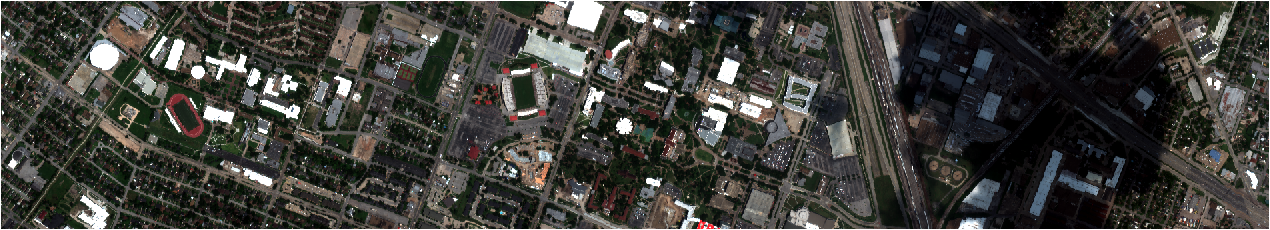}\\
(b) \includegraphics[width=.95\linewidth]{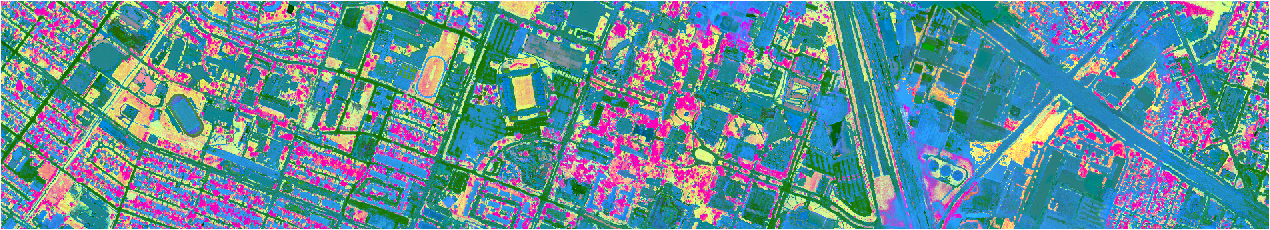}\\
(c) \includegraphics[width=.95\linewidth]{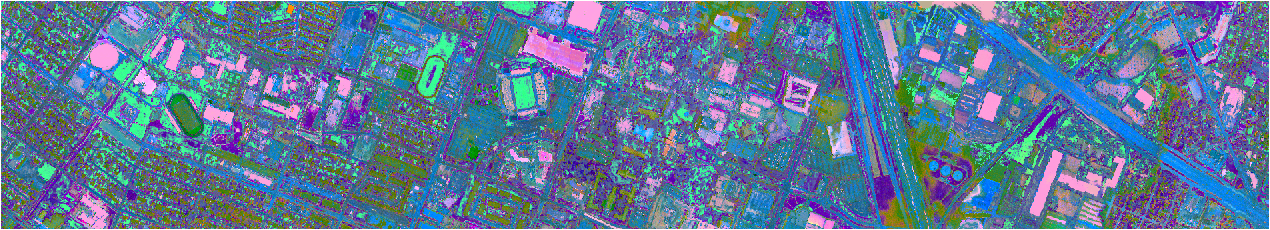}\\
(d) \includegraphics[width=.95\linewidth]{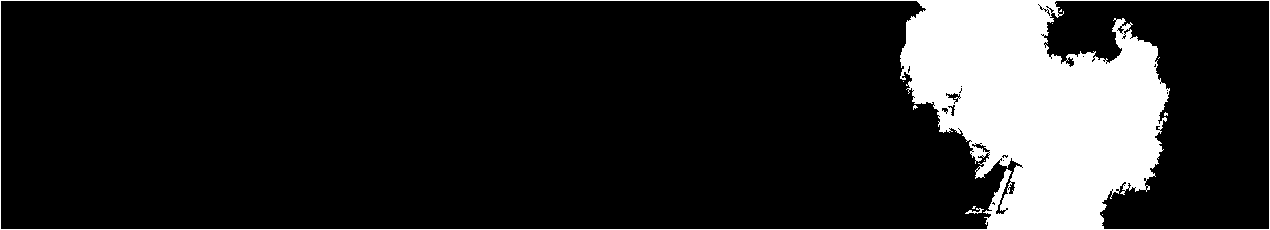}\\
\caption{Domains reprojected by KEMA. (a): original CASI image. (b): first three dimensions of the latent space (R: 1, G: 2, B: 3). (c): dimensions 4-6. (d): cloud mask defining the two domains. \label{fig:visuProjHouston}}
\end{figure}

\begin{figure*}[ht]
\begin{tabular}{p{.7cm}cc}
& HSI & HSI+LiDAR+AVG+MV \\
Raw & \includegraphics[width=.45\linewidth]{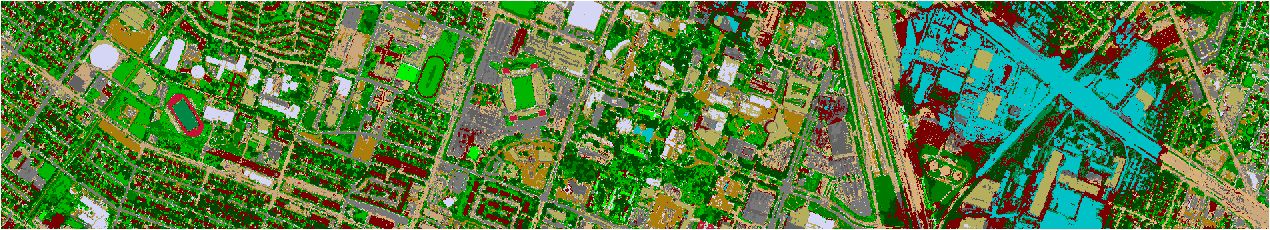} & \includegraphics[width=.45\linewidth]{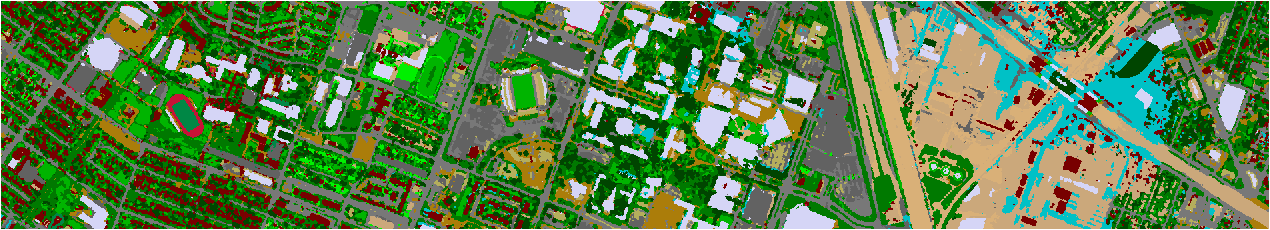} \\
HM & \includegraphics[width=.45\linewidth]{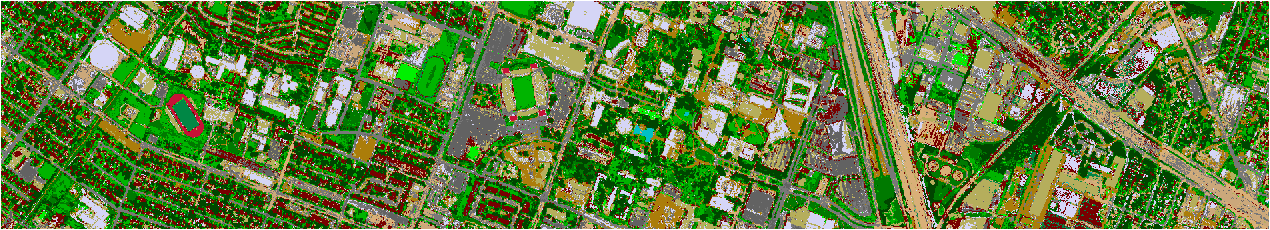} & \includegraphics[width=.45\linewidth]{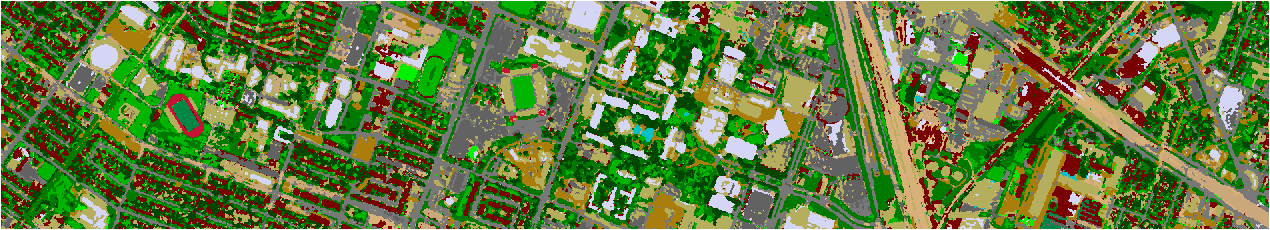} \\
KEMA & \includegraphics[width=.45\linewidth]{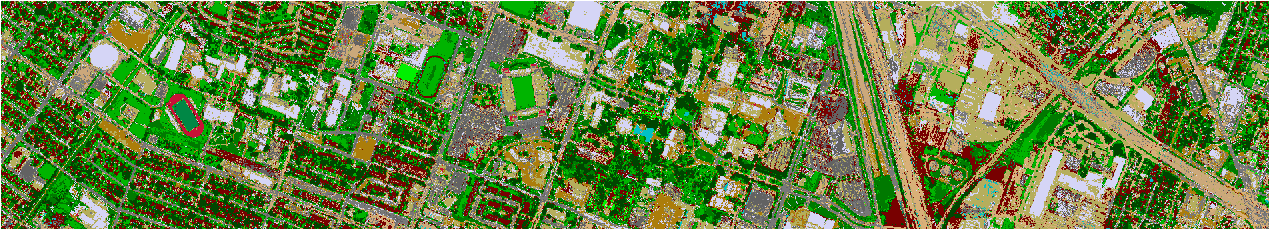} & \includegraphics[width=.45\linewidth]{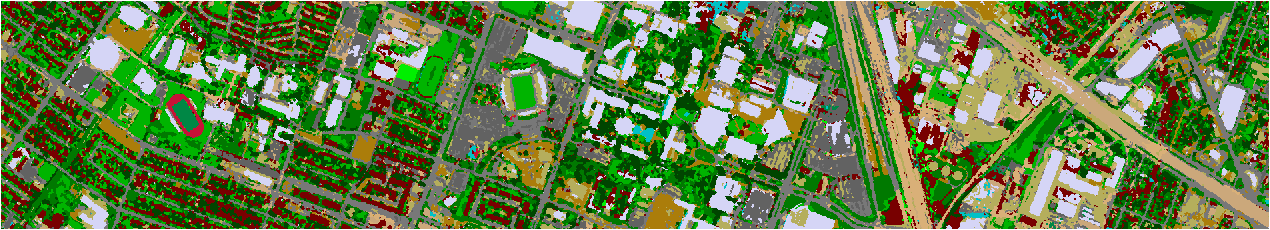} \\
 
Test  & \includegraphics[width=.45\linewidth]{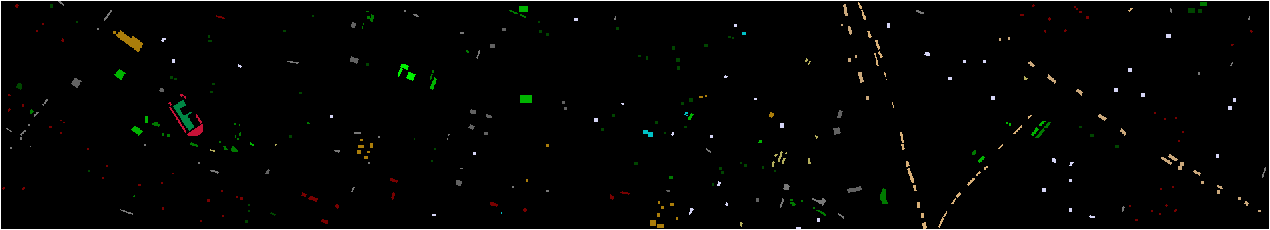} & \includegraphics[width=.45\linewidth]{./figs/figsHouston/mask.png}\\
\end{tabular}
\caption{Classification maps for the three settings (Raw, HM and KEMA). (left) using the spectral bands; (right) performing a majority voting on the map obtained by staking HSI, LiDAR and AVG features (for averaged numerical results, see Tab.~\ref{tab:svmrbf_houston}). Bottom line shows the test samples and the cloud mask. \label{fig:mapsHouston}}
\end{figure*}

In this experiment, we aim at compensating the reduction in reflectance due to a shadow casted by a large cloud. We consider a hyperspectral image acquired by the CASI sensor over Houston (see Fig.~\ref{fig:visuProjHouston}a) ). The data were originally provided to the community for the data fusion contest 2013~\cite{Deb14}\footnote{The data can be found at \url{http://www.grss-ieee.org/community/technical-committees/data-fusion/}}. The contest was framed as a land use classification contest, where 15 land use classes were to be detected using two data sources: the hyperspectral image mentioned and a LiDAR DSM. The specificity of the contest is that the test pixels are partly located under a shadow cast by clouds (see Fig.~\ref{fig:mapsHouston}d), thus raising the need for compensation algorithms. In our analysis, we compare three strategies for handling the hyperspectral image: using it without further processing (`Raw'), applying a histogram matching (HM) on the shadowed area (the strategy also used before extracting features in~\cite{Tui15}) and the proposed KEMA aligning the pixels under the shadow and those illuminated. For both the HM and KEMA, we define the shadowed pixels by defining a cloud mask by thresholding band 130 and then applying morphological operators to remove salt and pepper noise within the bigger connected component representing the shadow (cf. the mask in Fig.~\ref{fig:visuProjHouston}d).

In this experiment, we align the dataset using 20 labeled pixels per class. We use only classes occurring in both domains (shadowed and illuminated). Additionally, we sample randomly 200 unlabeled pixels per class. As for the first example, the kernel used in KEMA is an RBF with $\sigma_m$ bandwidth estimated as half of the median distance between the points of the domain. This is very important in this experiment, since it allows to have a much narrower bandwidth for the kernel acting on the shadowed domain than the one used in the illuminated domain. We classify using a support vector machine with RBF kernel, whose parameters are found by cross validation ($\sigma \in [0.01,0.1]$, $C \in [1,100]$). We train the classifier on 95\% of the training set available and predict on two validation datasets: the entire test set and the test samples under the shadowed area. We consider three feature sets, as detailed in Table~\ref{tab:fsHouston}, and use them in three experiments: the first using only the HSI, the second adding LiDAR-derived features, and the third adding contextual features extracted from the optical bands. A last setting, called MV, uses all features, and also applies a majority voting on the solution. The experiments are repeated 10 times by varying the labeled pixels in KEMA and those picked for classification: therefore we report the average and standard deviation.

\begin{table}[h!]
\caption{Three feature types used in the experiment. Number in brackets is the number of features involved in each group.\label{tab:fsHouston}}
\begin{tabular}{cp{3cm}|p{3.5cm}}
\hline
 & Raw / HM & KEMA \\
\hline\hline
HSI & Hyperspectral bands (144) & KEMA aligned features (50)\\\hline
 LiDAR & \multicolumn{2}{p{6.5cm}}{LiDAR band + opening and closing by reconstruction features with convolution of size $[7, 19 ,31]$ pixels (7)} \\\hline
 AVG & \multicolumn{2}{p{6.5cm}}{Average filters, window size $3$, applied on the: }\\
 & 10 first principal component projections (10) & 10 first KEMA projections (10) \\\hline
\end{tabular}
\end{table}

\begin{figure}[!t]
\begin{tabular}{cc}
 \includegraphics[width=.45\linewidth]{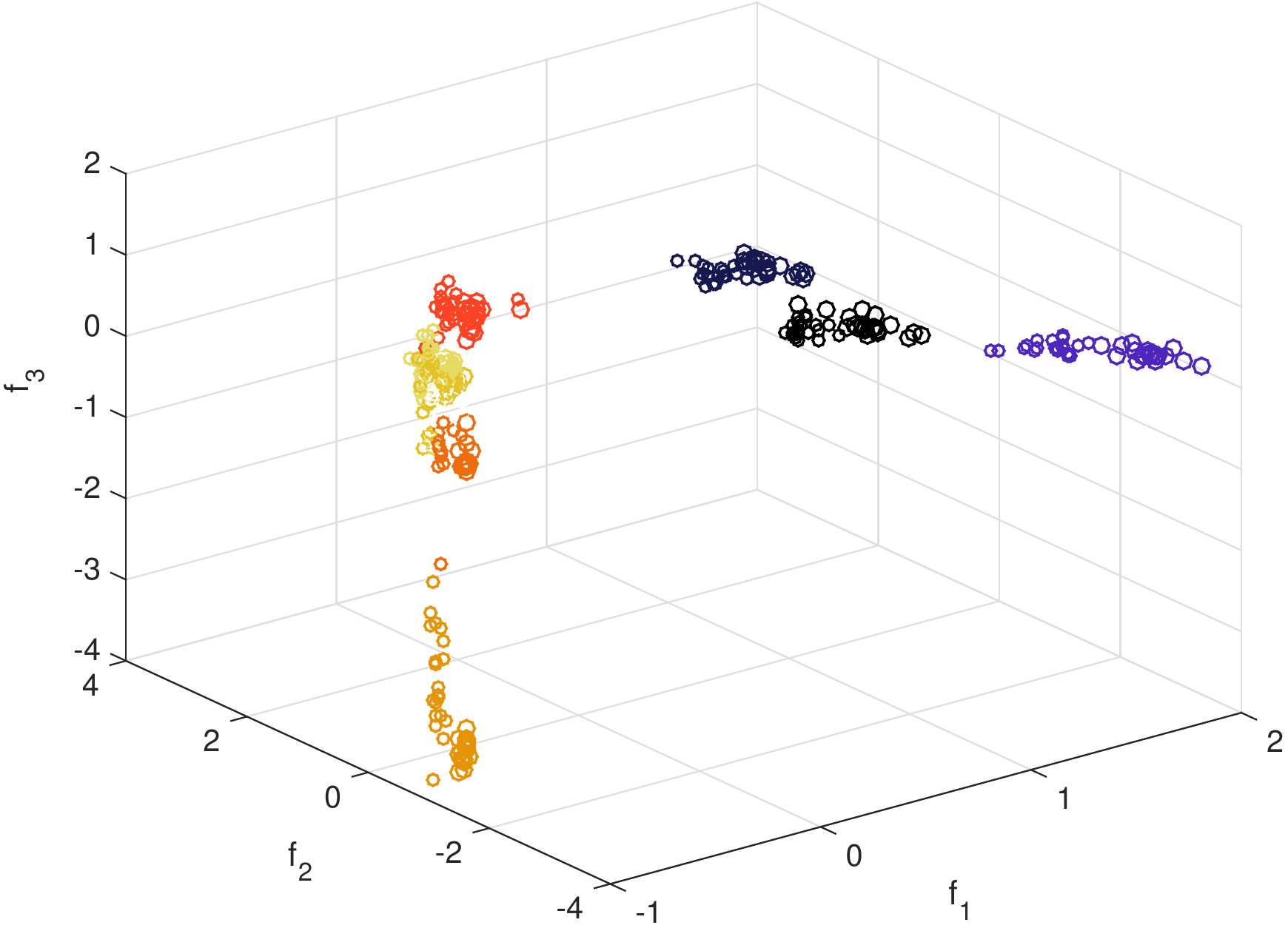}&
 \includegraphics[width=.45\linewidth]{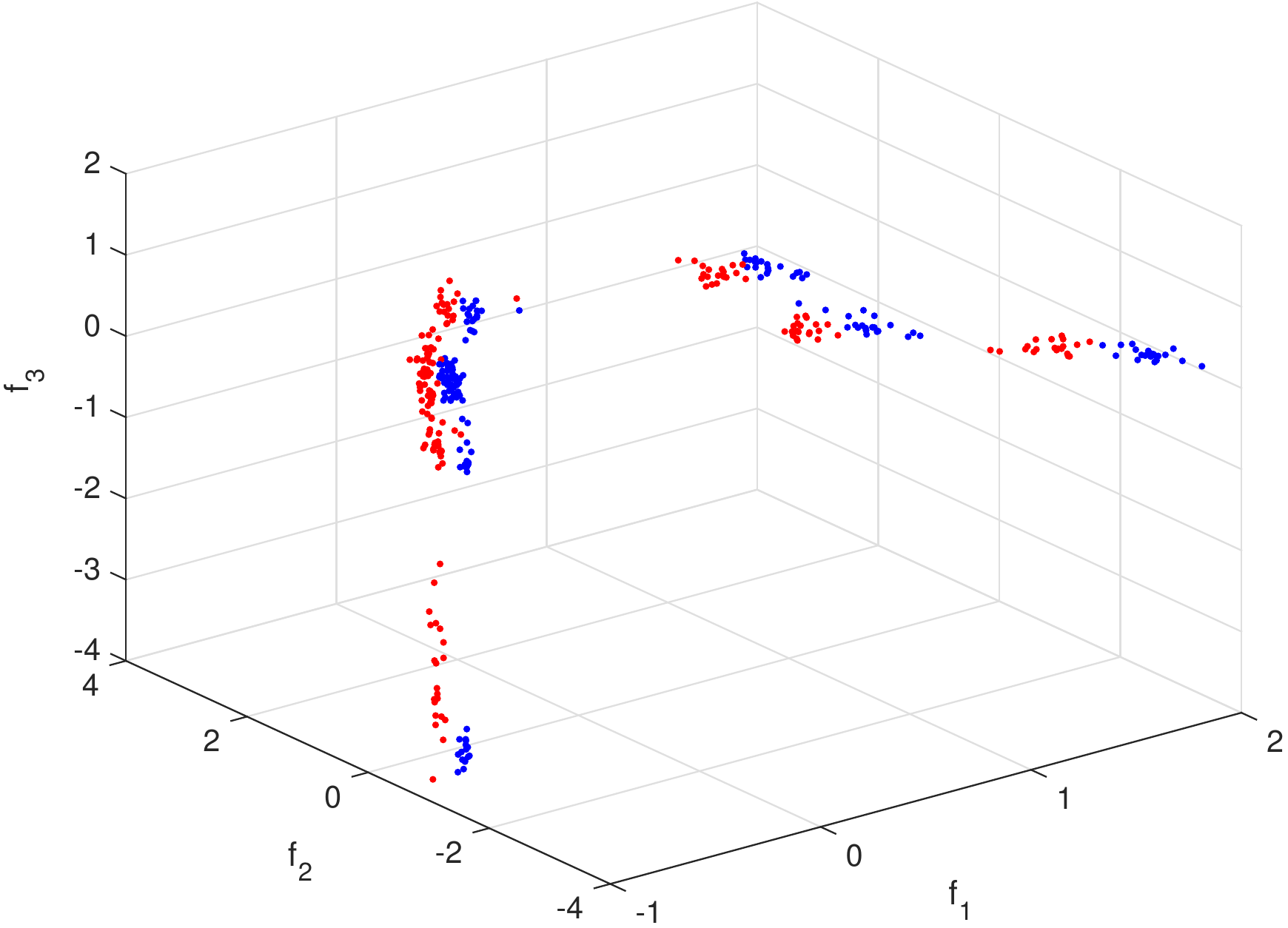}\\
 (a) per class & (b) per domain \\
\end{tabular}
\caption{Projection per class (a) and per domain (b, shadow is in blue and illuminated in red) for the Houston data. \label{fig:fs}}
\end{figure}

The projections extracted by KEMA are visualized in Fig.~\ref{fig:visuProjHouston} (geographical space, for projections $[1-3]$ and $[4-6]$) and Fig.~\ref{fig:fs} (feature space for dimensions $[1-3]$). At a first glance, the aligned features seem to be less dependent on the presence of the shadow than the original image (some artifacts remain at the border, due to the binary nature of the cloud mask). This is confirmed in the feature space, where the two domain seem correctly aligned both in terms of classes and domains.

The classification results reported in Table~\ref{tab:svmrbf_houston} confirm these intuitions: KEMA is able to provide higher classification performance by working in the aligned latent space. The use of the raw images (`Raw' column), even though satisfactory on the global test set (OA of 85.5\% in the best case), completely fails under the shadowed area (best OA: 23.8\%). This can be also appreciated in the classification maps (first row in Fig.~\ref{fig:mapsHouston}): from the maps it is clear that the shadow drains most of the shadowed pixels in the class `water' (in cyan). Even including LiDAR features (right column of Fig.~\ref{fig:mapsHouston}) does not solve entirely the problem and basically shifts most of the shadowed pixels in the class `highway' (in beige). Using HM improves drastically the solution under the shadow, since the accuracy goes from 23.8\% to 75.1\% on average. Histogram matching solves the problem globally and provides the scaling and centering of the histogram necessary to make the images more similar, but still fails at accounting for subtle local variations, thus still leading to heavy misclassifications in the final map, in particular the highway being classified as buildings (see second row of Fig.~\ref{fig:mapsHouston}). Finally, KEMA solves the problem locally by the flexibility of the kernel mapping: the accuracies are the highest (also matching those of the winners of the contest, who created an entirely \emph{ad-hoc} system for this specific image) and reach an average of 94.3\%, but also show an almost identical performance in the shadowed area (91.5\%). The alignment has made the two domains more similar and the mismatch between domains becomes almost invisible in the classification maps (third row of Fig.~\ref{fig:mapsHouston}).

\begin{table}[h!]
\caption{Classification results (Overall accuracy, in \%) for the Houston data. \label{tab:svmrbf_houston}}
\setlength{\tabcolsep}{7pt}
\begin{center}
\begin{tabular}{cp{1.1cm}|c|c|c}
\multicolumn{5}{c}{Entire test set} \\\hline
\multicolumn{2}{c}{HSI processing:}& Raw &  HM & KEMA (us)  \\ 
\hline
HSI && 71.0 $\pm$ 0.1 & 79.5 $\pm$ 0.4 &83.8 $\pm$ 1.9\\
$\hookrightarrow$&+ LiDAR& 83.4 $\pm$ 0.2 &86.4 $\pm$ 0.7& 89.4 $\pm$ 1.4   \\
{ } $\hookrightarrow$& + AVG & 85.1 $\pm$ 0.2 &84.5 $\pm$ 0.4& 93.0 $\pm$ 0.8   \\
{ } { } $\hookrightarrow$ &+ MV & 85.5 $\pm$ 0.2 & 86.0 $\pm$ 0.3& 94.3 $\pm$ 0.8  \\
\hline
\multicolumn{5}{c} {}\\
\multicolumn{5}{c}{Shadowed areas in the test set}\\
\hline
HSI &&  04.2 $\pm$ 0.1 & 67.4 $\pm$ 0.7 &70.0 $\pm$ 1.0  \\
$\hookrightarrow$&+ LiDAR& 22.5 $\pm$ 0.3 &77.1 $\pm$ 1.3& 82.6 $\pm$ 5.4  \\
{ } $\hookrightarrow$& + AVG &  23.2 $\pm$ 1.2 &73.6 $\pm$ 0.8& 90.4 $\pm$ 4.9  \\
{ } { } $\hookrightarrow$ &+ MV &  23.8 $\pm$ 1.2 &  75.1 $\pm$ 0.9&91.5 $\pm$ 4.5  \\
\hline
\hline
\end{tabular}
\end{center}
\end{table}

\subsection{Multi-source image classification without labels}\label{sec:wesma}

In the last experiment, we break the requirement for labeled data in all domains. To do so, we need to reduce the flexibility of KEMA by adding a requirement on \emph{partial spatial overlap between the scenes}. This can be understood as follows: KEMA is a spectral registration method that uses the labels as anchor points (or \emph{ties}) to register the domains spectrally. If one of the domains is unlabeled, it is not possible to register them, since the $\L_s$ and $\L_d$ matrices in Eq.~\eqref{primalKEMA} cannot be computed. As a consequence, we can only preserve the inner domain geometry using $\L_g$, but there is no way to find the matching between domains. 

\begin{figure}
\includegraphics[width=\linewidth]{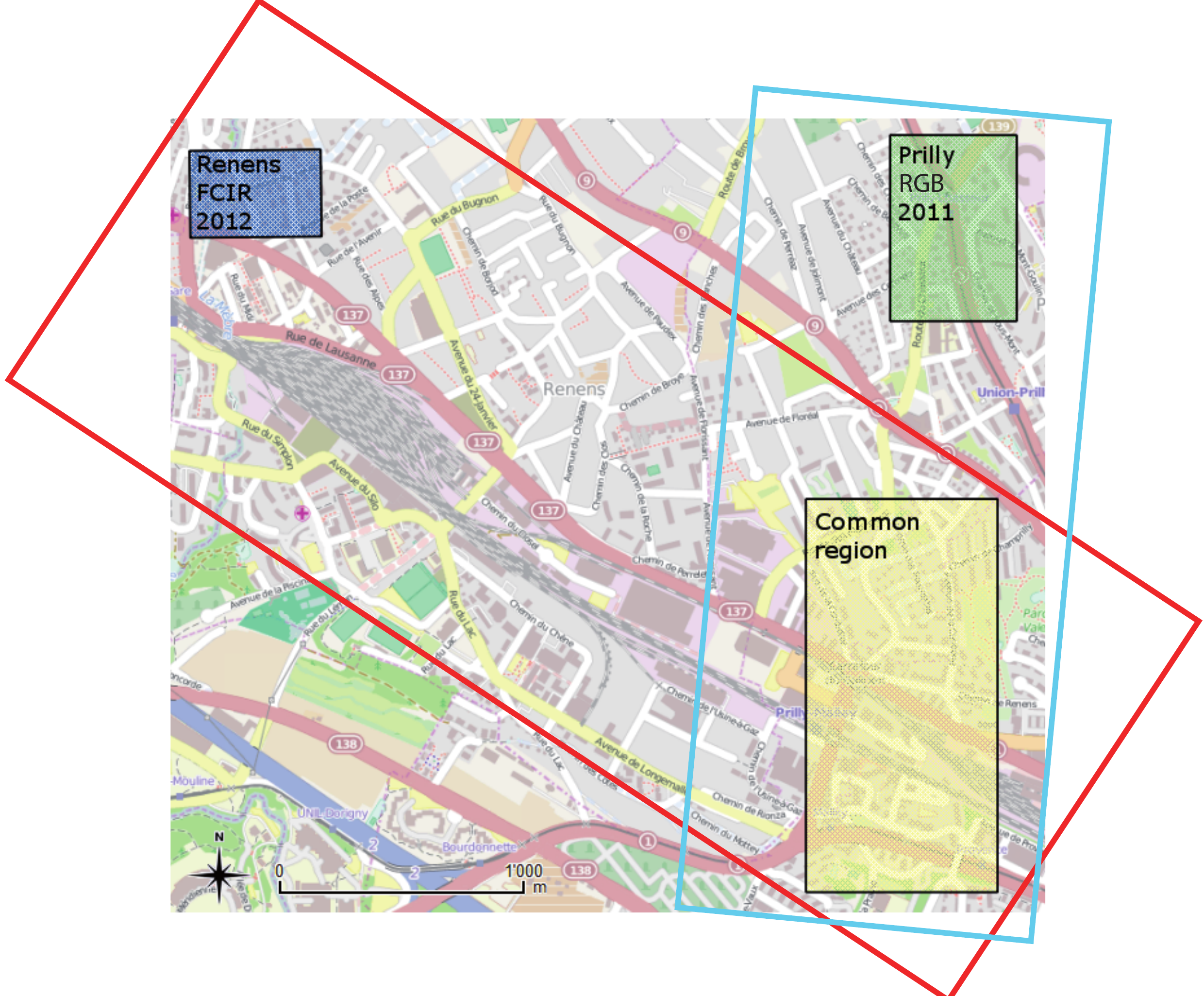}
\caption{Setting of the multi-source experiment. The cyan square represents the source domain image (RGB) and the red square the target domain image (NIR-R-G). They share a spatial subset, where the semantic ties are used to align the domains. The dark blue, green and yellow square are the image detailed in Fig.~\ref{fig:wesmaimgs}, used for both the semantic ties definition  and the numerical assessment.
\label{fig:wesmadata}}
\end{figure}

\begin{figure*}
\setlength{\tabcolsep}{3pt}
\begin{tabular}{cc|cc|c}
\includegraphics[height=3.8cm]{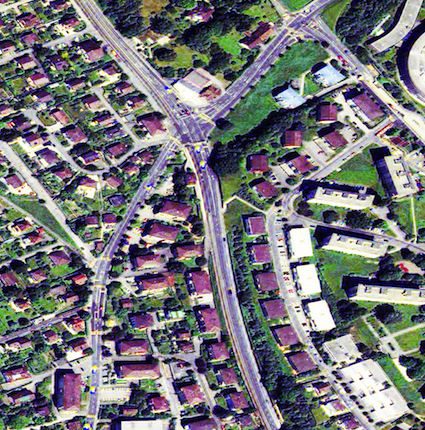}&
\includegraphics[height=3.8cm]{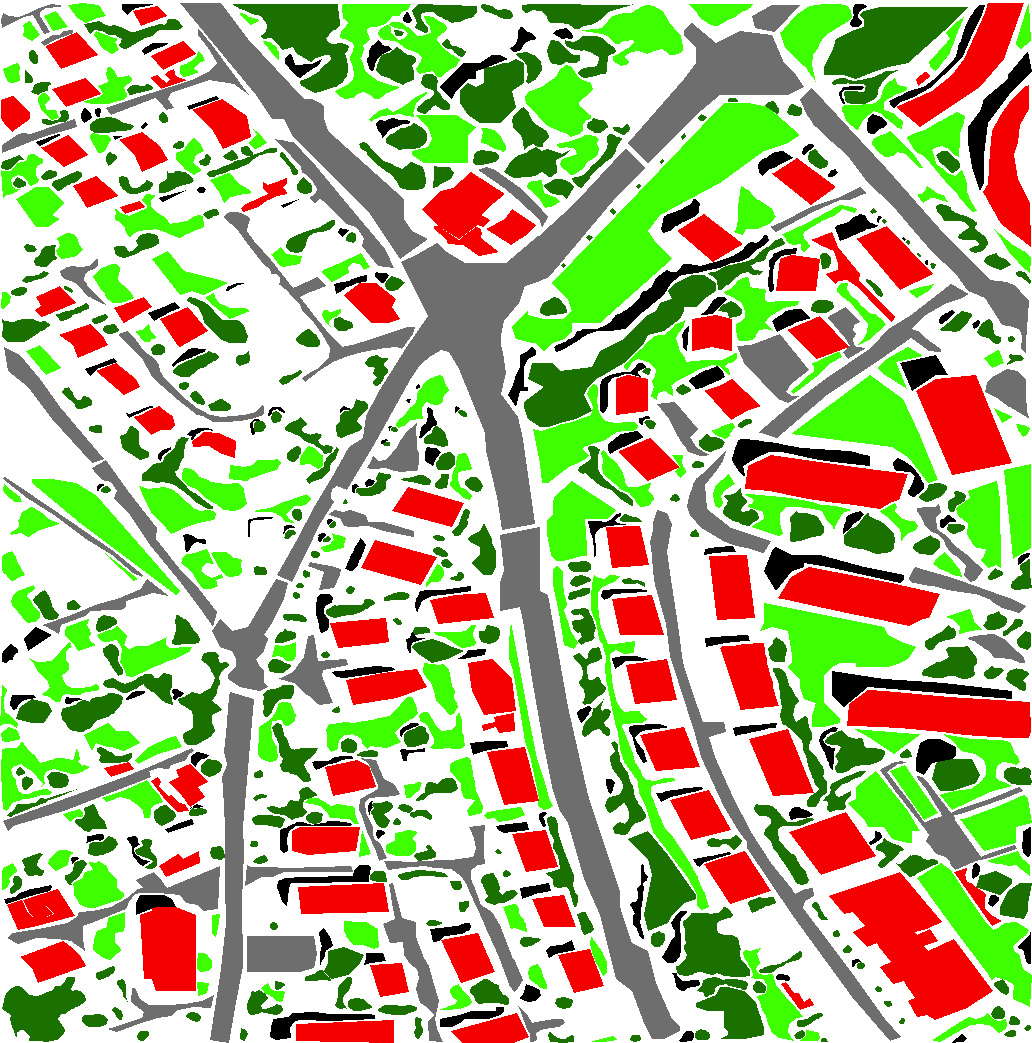}&
\includegraphics[height=3.8cm]{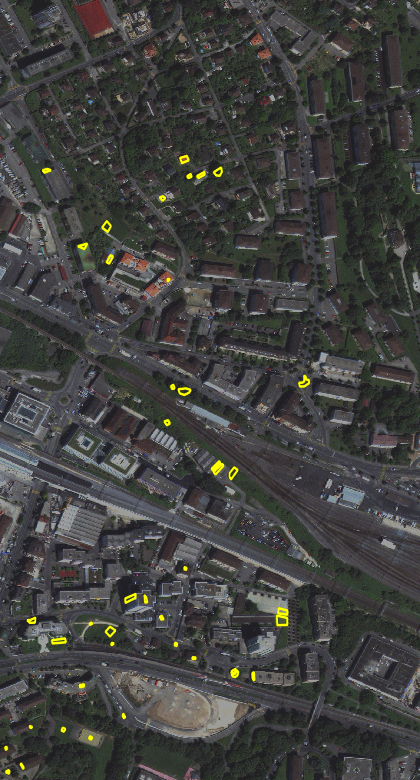}&
\includegraphics[height=3.8cm]{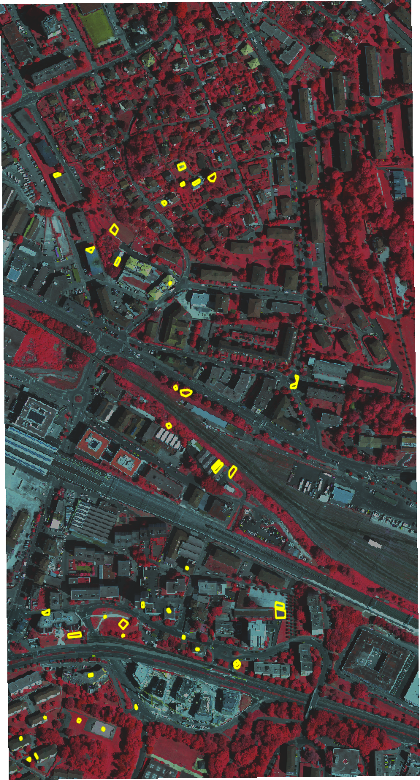}&
\includegraphics[height=3.8cm]{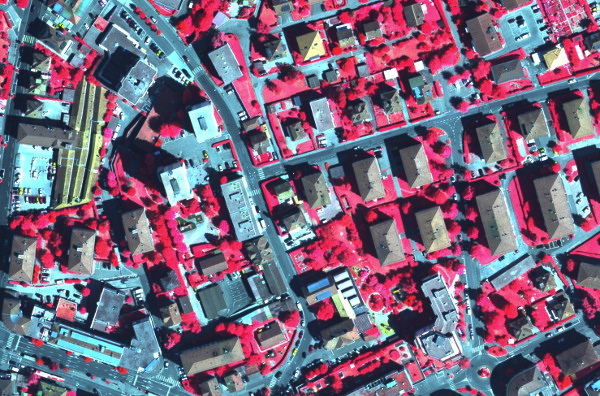}\\
\multicolumn{2}{c|}{Prilly: source domain} & \multicolumn{2}{c|}{Spatially overlapping area} &  Renens: target domain (unlabeled) \\
\multicolumn{2}{c|}{(RGB)} & \multicolumn{2}{c|}{with semantic ties} &  (NIR-R-G) \\
\multicolumn{2}{c|}{(a)} & \multicolumn{2}{c|}{(b)} &  (c) \\
\end{tabular}
\caption{Images involved in the multi-source experiment (corresponding to the dark blue, green and yellow squares in Fig.~\ref{fig:wesmadata}).\label{fig:wesmaimgs}}
\end{figure*}

When using geographical data (as remote sensing data), a special case can break this requirement: whenever the domains are (at least partially) co-located in space. In this case, represented in Fig.~\ref{fig:wesmadata}, the two images share a spatial region, where we can co-locate objects, for instance by feature keypoint matching or by manual registration. Once these matches are found, they can be used to build the matrix $\L_s$, since, even if we ignore their class, we know that the pixels of the objects matched belong to the same class (they are known as \emph{semantic ties}~\cite{Mon13}). This type of weakly supervised alignment has been recently proposed in~\cite{Mar16} and we use it here prior to aligning the data spaces with KEMA. 
The experiment is set as follows: 
\begin{itemize}
\item[-] We use an RGB image (0.6m resolution) over the area of Prilly, a neighbourhood of Lausanne, Switzerland as source domain. The area is labeled into five classes (roads, buildings, trees, grass and shadows) by manual photo-interpretation, see Fig.~\ref{fig:wesmaimgs}a.
\item[-] An FCIR (false colour infrared with NIR-G-B bands) ortho-photo of the area of Renens (another neighbourhood of Lausanne), at 0.25 cm resolution, is used as target domain and the labels are this time kept hidden (they are only used for validation), see Fig.~\ref{fig:wesmaimgs}c.
\item[-] To find the projections with KEMA, we use an overlapping area between the two images. The overlapping areas are not registered nor they are at the same spatial resolution: to match them, we provide 40 tie object by manual drawing in both images (the operation takes less than 5 minutes), see Fig.~\ref{fig:wesmaimgs}b.
\end{itemize}

We use the labels in the source and the semantic ties to construct the $\L_s$ matrix. For the $\L_d$ matrix, we extracted the graph Laplacian from a dissimilarity matrix with values $1$ for pixels from different classes in the source and $0.5$ when issued from different objects in the semantic ties. We give a smaller penalization in the latter case, since two pixels coming from different objects can still belong to the same class. Once the domains are aligned, we train a linear SVM with 100 labeled pixels per class from the source domain (the RGB image) and test 400 pixels per class in the target domain (the FCIR image). 

\begin{figure}[h!]
\textcolor[rgb]{1,1,1}{a}\hspace{1.2cm} Unprojected \hspace{2.5cm} KEMA\\
\includegraphics[width=\linewidth]{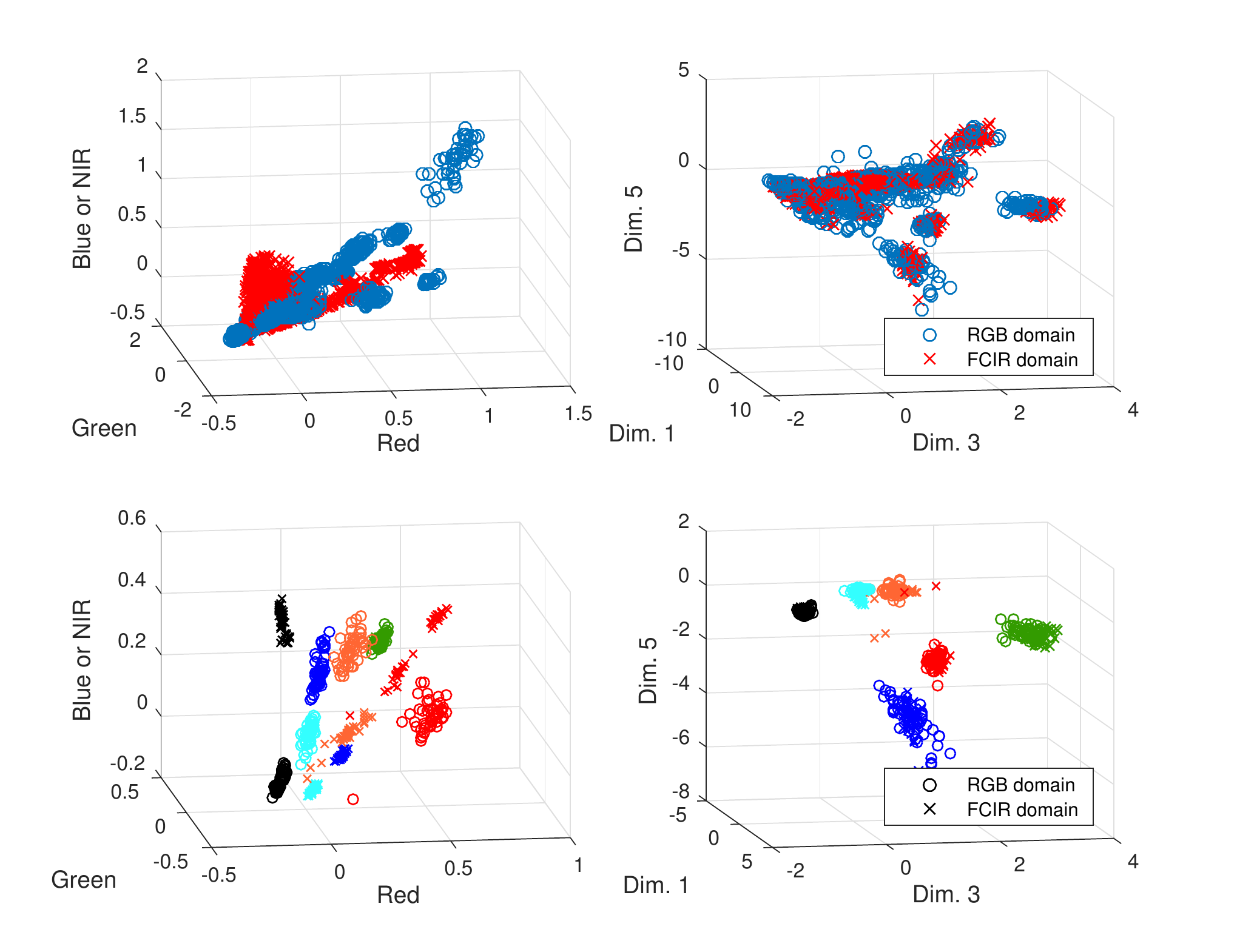}
\caption{Projections found by KEMA, colored by domain (top) and by object in the semantic ties set (bottom, six objects shown). The left panel shows the unprojected data [x axis: R, y axis: G, z axis: NIR or B], the right panel shows the projections by KEMA [Projections 1, 3 and 5].   \label{fig:wesmaproj}}
\end{figure}

\begin{figure}[h!]
\includegraphics[width=\linewidth]{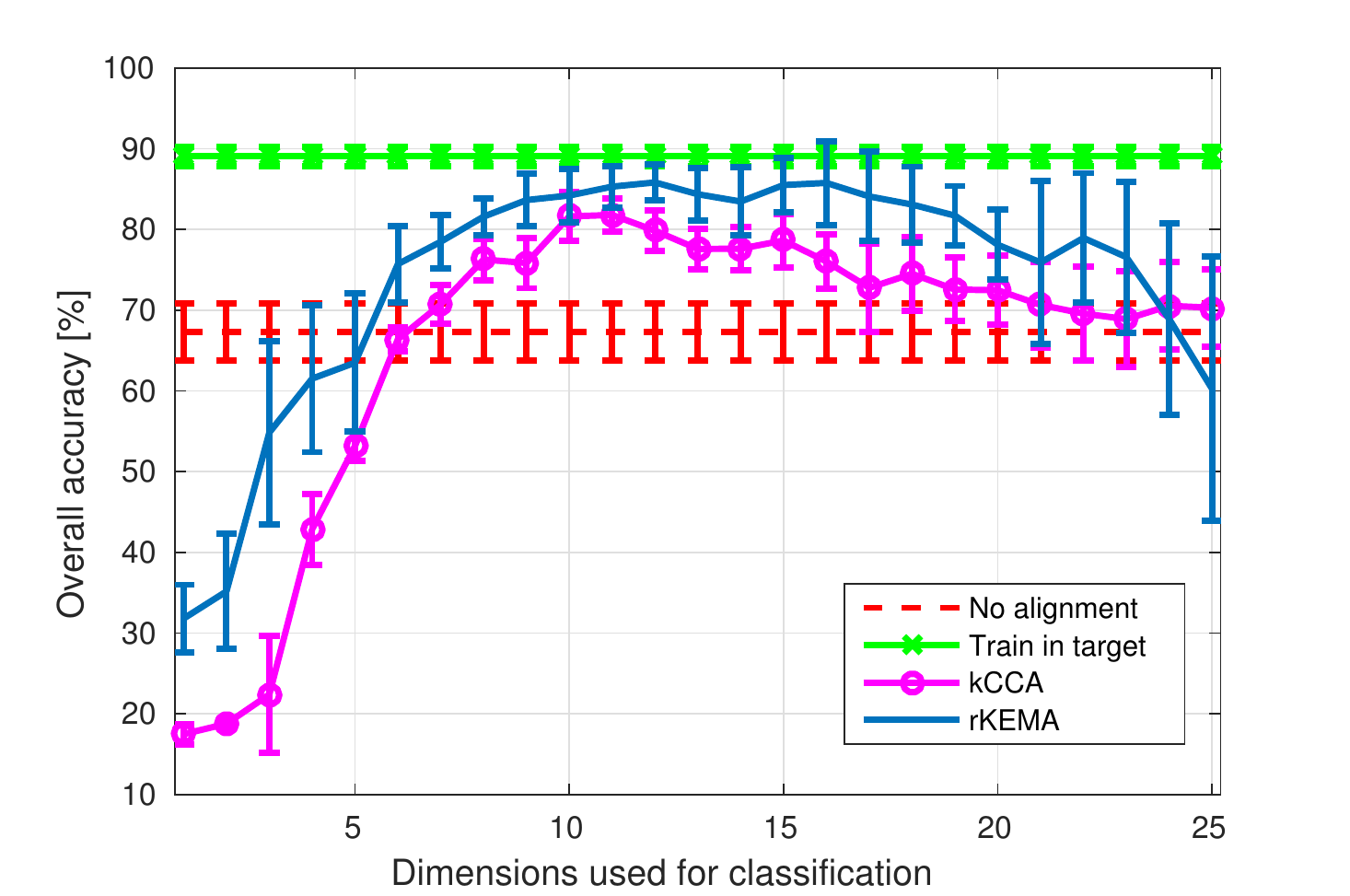}
\caption{Classification performances by a linear SVM using the labeled samples from the source domain (RGB) as they are (red line) or projected by KEMA (blue line) or kCCA (magenta line). In green a baseline obtained by training with labeled pixels form the target domain (FCIR).  \label{fig:wesmacurves}}
\end{figure}

The projections retrieved are illustrated in Fig.~\ref{fig:wesmaproj}: as for the previous examples, KEMA shows aligned data spaces, but also discriminative in terms of objects aligned: the bottom line in Fig.~\ref{fig:wesmaproj} illustrates six objects among the 40 semantic ties used to find the alignment. Figure~\ref{fig:wesmacurves} reports the classification performance in the FCIR domain: starting with six dimensions, KEMA outperforms the case where the RGB image is used to predict the FCIR one without any adaptation\footnote{To maximize the performance of the `no alignment' case, we use the bands that share comparable wavelengths across domains: $X^s = [R, G]$, $X^t = [R, G]$.}: when using 13 dimensions, KEMA performs comparably to a model trained on labeled pixels form the target domain itself (green line in the figure). We compare these results to those obtained by applying kCCA~\cite{Lai00}. \diego{}{In order to compute the projection, we considered each object (each semantic tie in Fig.~\ref{fig:wesmaimgs}b) as a sample and used the spectrum of the most similar pixel to the object average to describe it. We then extract the kCCA projections between the 40 pairs of corresponding objects across image acquisitions. Back to the numerical results in Fig.~\ref{fig:wesmacurves},} the performance of KEMA is consistently better than that of kCCA. This is probably due to two reasons: 1) the fact that KEMA doesn't need a one-to-one correspondence and thus all the pixels in an object are taken into account for the projection and 2) that class separability is explicitly taken into account by using the labels in the source domain.

\section{Conclusions}\label{sec:conc}

In this paper, we presented a manifold alignment method based on kernels. The presented KEMA method is a feature extractor that finds projections from all the available source domains into a joint {\em latent} space, where data is semantically aligned and class separability enhanced. Compared to recent manifold alignment methods, KEMA offers a more flexible framework, going beyond simple linear transformations (scalings and rotations) of the input data. KEMA exploits a few labeled samples (or semantic ties) in each domain along with the wealth of unlabeled samples. KEMA reduces to solving a simple generalized eigenvalue problem, and has very few (and interpretable) hyperparameters to tune. We successfully tested KEMA in multi-temporal and multi-source very high resolution classification tasks, as well as on the task of making a model invariant to shadows for hyperspectral imaging. 

KEMA can be seen as a multivariate method for data pre-processing in general applications where multi-sensor, multi-modal, sensory data is acquired. The generality of the approach opens a wide field in remote sensing data processing applications. 
Our next steps with KEMA involve 1) performing semi-automatic atmospheric compensation in multi-temporal settings, 2) reduce the impact of the few labeled examples needed to perform the alignment, and 3) extend KEMA for challenging regression problems.

\section*{Acknowledgements}

The authors would like to thank the Hyperspectral Image Analysis group and the NSF Funded Center for Airborne Laser Mapping (NCALM) at the University of Houston for providing the hyperspectral data used in the shadow correction experiment, and the IEEE GRSS Data Fusion Technical Committee for organizing the 2013 Data Fusion Contest. They also would like to thank Swisstopo (\url{www.swisstopo.admin.ch}) for making available the FCIR orthophotos for academic use.

%\bibliographystyle{IEEEbib}
%\bibliography{refURBAN,align}

\end{document}